\documentclass[12pt]{iopart}

\usepackage{mathpazo}
\usepackage{color,cancel,wrapfig,lscape,array}
\usepackage{textcomp}
\usepackage{times}
\usepackage{ragged2e}
\usepackage{mathptmx}
\usepackage{hhline}
\usepackage{enumerate}
\usepackage{bbm}
\usepackage{units}
\usepackage{nicefrac}
\usepackage{graphicx}
\usepackage[dvips]{rotating}
\usepackage{color}
\usepackage{subfigure}
\definecolor{violet}{cmyk}{0.94,0.86, 0.04,0.08}
\usepackage[
colorlinks, 
linktocpage, 
bookmarksnumbered, 
bookmarksopen, 
pdfstartview=FitH, 
linkcolor=blue, 
citecolor=blue, 
urlcolor=magenta, 
filecolor=blue %
backref 
]{hyperref}

\usepackage{iopams}
\expandafter\let\csname equation*\endcsname\relax
\expandafter\let\csname endequation*\endcsname\relax
\usepackage{amsmath,amsfonts,amssymb,dsfont,amsxtra,amsopn,amstext,amsbsy}
\usepackage{amssymb,amsthm,amsthm,amsfonts}
\usepackage[T1]{fontenc}
\usepackage[latin1]{inputenc}

\usepackage{xr}

\newcommand{\ndg}{{\phantom{\dagger}}}
\newcommand{\dg}{\dagger}
\newcommand{\ew}[1]{\pmb{\big\langle} #1 \pmb{\big\rangle}}

\newcommand{\ket}[1]{| #1 \rangle}
\newcommand{\bra}[1]{\langle #1 |}
\newcommand{\bracket}[2]{\langle#1|#2\rangle}
\newcommand{\ketbra}[2]{\left|#1\right\rangle\hspace{-1.1mm}\left\langle #2 \right|}

\newcommand{\minus}{\hspace{-0.1cm}-\hspace{-0.1cm}}
\newcommand {\hc}{\text{h.c.}}

\begin{document}

\title{Opto-Nanomechanics Strongly Coupled to a Rydberg Superatom: Coherent vs. Incoherent Dynamics}

\author{Alexander Carmele, Berit Vogell, Kai Stannigel, and Peter Zoller}
\address{Institute for Theoretical Physics, University of Innsbruck,  and \\
Institute for Quantum Optics and Quantum Information, Austrian Academy of Sciences, Innsbruck, Austria}
\ead{alexander.carmele@uibk.ac.at}

\begin{abstract} 
We propose a hybrid optomechanical quantum system consisting of a moving membrane strongly coupled
to an ensemble of N atoms with a Rydberg state.
Due to the strong van-der-Waals interaction between the atoms, the ensemble forms an effective two-level system, a Rydberg superatom,
with a collectively enhanced atom-light coupling. 
Using this superatom imposed collective enhancement
strong coupling between membrane and superatom is feasible for parameters within the range of current experiments.
The quantum interface to couple the membrane and the superatom can be a pumped single mode cavity, or 
a laser field in free space, where the Rydberg superatom and the membrane are spatially separated.
In addition to the coherent dynamics, we study in detail the impact of the typical dissipation processes, in particular the radiative decay as a source for incoherent superpositions of atomic excitations. 
We identify the conditions to suppress these incoherent dynamics and thereby a parameter regime for strong coupling.
The Rydberg superatom in this hybrid system serves as a toolbox for the nanomechanical resonator allowing for a wide range of applications such as state transfer, sympathetic cooling and non-classical state preparation.
As an illustration, we show that a thermally occupied membrane can be prepared
in a non-classical state without the necessity of ground state cooling.
\end{abstract}

\section{Introduction}

The remarkable experimental developments of Cavity Optomechanics have essentially been based on linear coupling of the mechanical oscillator to the light field. This includes the preparation of mechanical resonators 
\cite{kippenberg_prl_microtoroid,gigan_nature_micromirror,jiang_optex_cavity_om,thompson_nature_membrane} in both classical and non-classical states via  sideband cooling, the observation of coherent  coupling with light\cite{o2010quantum,teufel2011sideband,chan2011laser}, and displacement detection at  the standard quantum limit \cite{naik2009towards,gavartin2012hybrid}. A main challenge in the field of opto-nanomechanics remains achieving non-linearities on the single phonon level, and strong coupling of the mechanical resonator to a two-level system in particular \cite{ramos_prl_defects}. Such nonlinearities are a key to generate entangled states and can be utilized for enhanced readout, quantum information processing and teleportation \cite{rabl_nature_transducers,stannigel_pra_transducers,marshall_prl_superpositions}.

In the present work we will study the coupling of a nanomechanical oscillator via light to a Rydberg superatom representing a two-level system. We will show that the strong coupling regime can be reached in this setup in the sense of a Jaynes-Cummings model \cite{chang_njp_cqed_mirrors}, where coherent couplings between the oscillator and the atom dominate dissipative effects. Our particular setup is motivated both by recent advances in experimentally realizing hybrid systems of nano-mechanical oscillators coupled to cold trapped atoms in experimentally compatible setups\cite{joeckel_apl_mechanical_dissipation,camerer_prl_om_cold_atoms,vogell_pra_cavity_enhancement} (see also \cite{hunger_crp_cold_atoms_om,hammerer_prl_single_atom,wallquist_pra,hammerer_pra_optical_lattices_om}), as well as the remarkable experimental achievements in realizing Rydberg superatoms with cold atomic ensembles \cite{jaksch_prl_rydberg,lukin_prl_rydberg,brennecke_nature_bec_cqed}. A Rydberg superatom consists of an ensemble of  $N$ cold atoms, which are  excited by light to the Rydberg state, where the dipole blockade mechanism based on the strong Van-der-Waals interaction between the Rydberg levels allows only a single collective excitation in the whole ensemble, thus forming an effective two-level system \cite{jaksch_prl_rydberg,lukin_prl_rydberg,brennecke_nature_bec_cqed}. The collective nature of the excitation representing the superatom leads to a collectively enhanced coupling to the light field of $\sqrt{N}$. In our proposal we exploit this enhanced coupling to achieve the strong coupling limit of Cavity Optomechanics  to the Rydberg superatom via  light field interactions. While the collective resonator - Rydberg superatom coupling benefits from the  $\sqrt{N}$ coupling, we show that dissipation (spontaneous emission of the atom)   can scale in an appropriate parameter regime as the single particle decay rate. 

The paper is  organized as follows. In Sec.~\ref{sec:overview} we will provide an overview introducing the two model systems of interest. We first describe the conceptually simple setup, where
both the mechanical resonator and the atomic ensemble are placed inside a single cavity, cf. Fig.~\ref{fig:1}(a),  and describe an alternative setup, where the mechanical resonator is coupled to a distant cloud of atoms (compare \cite{hunger_crp_cold_atoms_om,hammerer_prl_single_atom,wallquist_pra,hammerer_pra_optical_lattices_om}), cf. Fig.~\ref{fig:1}(c-d).
For the cavity mediated case, we discuss in detail the incoherent and coherent part of the dynamics in a microscopic model, cf. Sec.~\ref{sec:cavity_mediated}.
Writing the dynamics in a collective basis in Sec.~\ref{sec:collective_dynamics}, 
we discuss an effective model and show that   the
strong coupling regime for experimental accessible parameters can be reached. 
The alternative set-up, in which the mechanical resonator and the atomic ensemble is
spatially separated, is described in  Sec.~\ref{sec:laser_mediated}, where the derivation
is summarized.
Both setups allow to utilize the Rydberg superatom as a toolbox for optomechanical
experiments in the strong coupling regime, cf. Sec.~\ref{sec:toolbox}. 
We discuss sympathetic cooling, state transfer and the 
preparation of non-classical states for a membrane coupled to heat bath, before we conclude
the paper in Sec.~\ref{sec:conclusion}.
\section{Overview}
\label{sec:overview}
Before giving details of the proposed setup we summarize in this section
the important results and main features of the derivation 
to strongly couple nanomechanics to a Rydberg superatom.
The main goal is to realize a nonlinear interaction between a moving membrane
and a two-level system in the strong coupling regime. 
This has applications such as non-classical mechanical state preparation.
The coherent part of the dynamics is governed by a Jaynes-Cummings type of interaction:
\begin{eqnarray}
\label{eq:jcm}
H_{\rm JCM} = 
\hbar\omega_m b^\dg b
+
\hbar
\frac{\omega_{s}}{2}
\sigma_z
+
\hbar G_{\rm eff} \left( b \ \sigma_+ + b^\dg \ \sigma_-  \right),
\end{eqnarray}
where $b^{(\dg)}$ is the annihilation (creation) operator of the mechanical 
mode of the membrane with frequency $\omega_m$, 
and $\sigma_\pm,\sigma_z$ are the 
Pauli operators for a two-level system with frequency $\omega_s$.
$G_{\rm eff}$ is the coupling constant and governs the 
excitation transfer between the membrane and the two-level system.
In addition to the coherent dynamics we also have dissipative dynamics such as phonon heating and radiative decay of the excited atomic states.
The complete dynamics of the system is governed by the following 
master equation
\begin{eqnarray}
\label{eq:jcm_meq}
\partial_t \rho 
= 
-\frac{i}{\hbar} \left[H_{\rm JCM},\rho \right] 
+
\mathcal D [\sqrt{\gamma} \ b]
\rho 
+
\mathcal D [\sqrt{\Gamma} \sigma_-]
\rho .
\end{eqnarray}
Here, the decay of the mechanical mode $\gamma$ and the two level system $\Gamma$ are written in Lindblad form, i.e. 
$\mathcal D [ A ]
\rho := 2 A \rho A^\dg - \lbrace A^\dg A, \rho \rbrace $.
In order to reach the strong coupling regime, i.e. $G_{\rm eff} \gg \gamma,\Gamma$, 
we propose to use a Rydberg superatom as the two-level system
 since it provides a strongly enhanced atom-light coupling\cite{pritchard_review_rydberg,brennecke_nature_bec_cqed}. Here, the question arises whether the benefit of the enhanced coherent dynamics is diminished by also enhanced dissipative dynamics. 

A Rydberg superatom consists of an ensemble of $N$ atoms with highly excited
Rydberg states.
These Rydberg states interact via a strongly repulsive van-der-Waals potential. 
As a consequence, if a single Rydberg excitation is present, neighbouring Rydberg states are shifted
out of the laser resonance and further Rydberg excitations are suppressed
within a so-called Rydberg blockade radius \cite{jaksch_prl_rydberg,lukin_prl_rydberg,brennecke_nature_bec_cqed}.
We are interested in a situation where this radius is larger than the size of the atomic ensemble
to allow only a single Rydberg excitation. 
In this limit a strong non-linearity is created via the enhancement of the Rabi frequency
by the square root of the number of atoms in the ensemble.
For resonance conditions identified below, we find that only the coherent dynamics benefits from the collective enhancement and thereby outrivals the incoherent part. Hence, the atomic ensemble can be described as an 
effective two-level system \cite{honer_prl_artificial_atoms,brennecke_nature_bec_cqed}.

\begin{figure}[b!]
\centering
\includegraphics[width=0.9\textwidth]{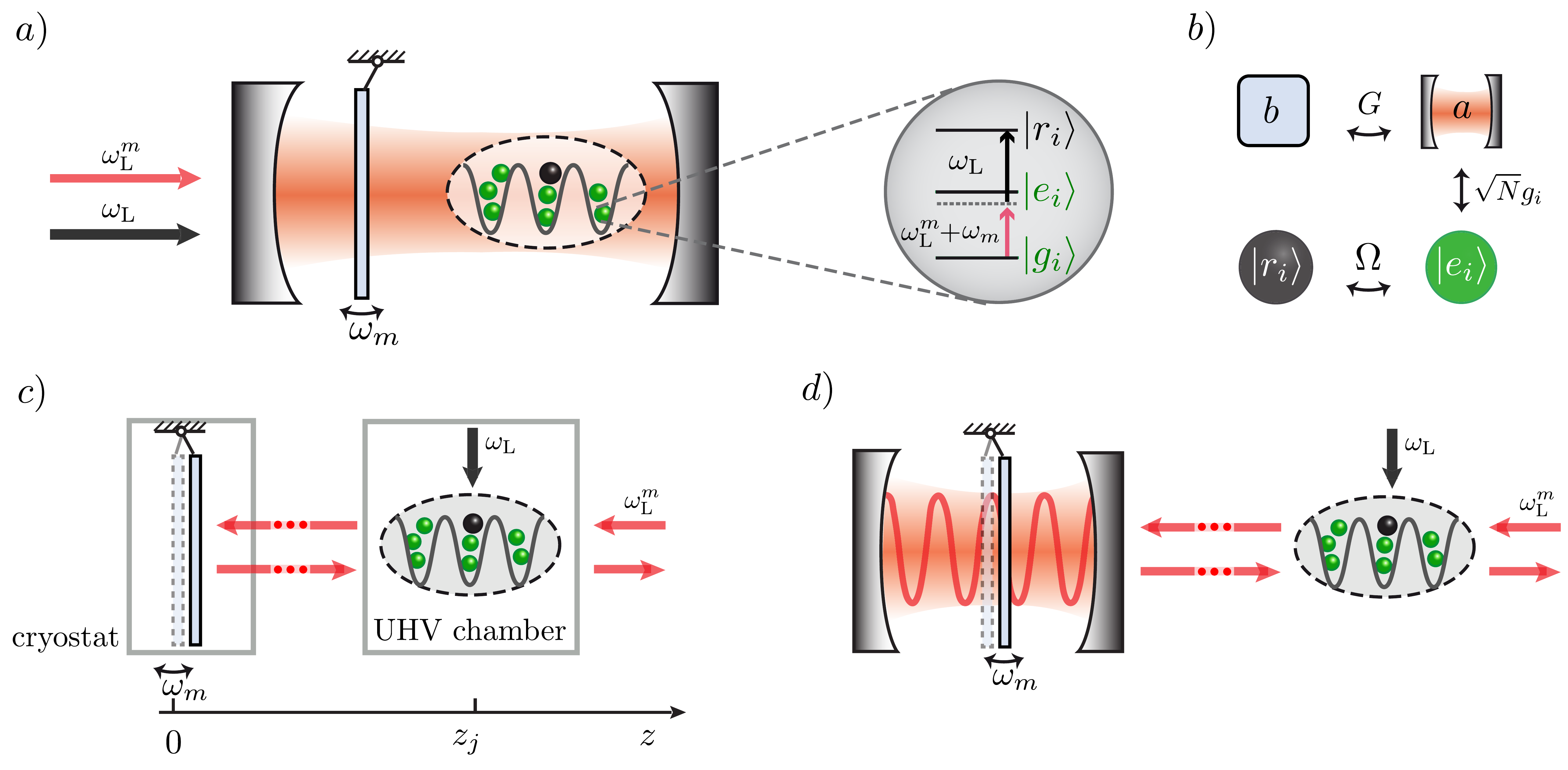}
\caption{An ensemble of three-level atoms coupled to a moving membrane. (a)
Due to the Rydberg blockade effect, a superatom is formed and allows strong coupling between the internal degrees of the superatom and the membrane. {\it Inset}: Internal states of atom $i$. (b) Coupling mechanism. Membrane couples via cavity to internal states of atomic ensemble. (c) Modular long-distance setup of a membrane inside a cavity coupled to an ensemble of Rydberg atoms. The membrane can be kept in a cryogenic environment, and the atoms at a distance in a vacuum chamber, cf.\cite{hammerer_pra_optical_lattices_om}. (d) Advancement: cavity-enhanced long-distance coupling, cf. \cite{vogell_pra_cavity_enhancement}.}
\label{fig:1}
\end{figure}

The setup we have in mind can be implemented in two different ways. One possibility is to consider both the mechanical oscillator and the Rydberg superatom inside a high finesse cavity that mediates the interactions, cf. Sec. \ref{sec:cavity_mediated}
and \ref{sec:collective_dynamics} and Fig.~\ref{fig:1}(a). An alternative setup is discussed in Sec.~\ref{sec:laser_mediated} and depicted in Fig.~\ref{fig:1}(c-d). It consists of a mechanical oscillator inside a cavity [Harris paper] coupled via a mediating laser to a distant atomic Rydberg ensemble.
This proposal thereby opens the possibility of strongly coupling a nanomechanical oscillator  to a Rydberg superatom, which then serves as a toolbox for optomechanical experiments in the non-classical regime, cf. Sec.~\ref{sec:toolbox} .

In the following we start with a microscopic model for the full system. Subsequently, we derive the effective Jaynes-Cummings type dynamics given in Eq.~\eqref{eq:jcm_meq}.

\section{Model} \label{sec:cavity_mediated}
%
In this section we discuss the details of the setup as depicted in Fig.~\ref{fig:1}(a).
It consists of a moving membrane, a high finesse cavity 
and an ensemble of $N$ atoms  (labelled $i$) with a Rydberg state $\ket{r_i}$.
We assume the Rydberg state is excited in a two-photon process from the ground $\ket{g_i}$ via an intermediate excited state 
$\ket{e_i}$ (see inset Fig.~\ref{fig:1}(a)) \cite{pritchard_review_rydberg,brennecke_nature_bec_cqed}.
The full Hamiltonian of the setup reads
\begin{eqnarray}
\label{eq:Hamiltonian}
H = H_0 + H_{\rm int},
\end{eqnarray}
where $H_{ \rm int}$ consists of the different
interaction Hamiltonians and $H_0$ is the free evolution 
given by $( \hbar =1)$:
\begin{eqnarray}
\label{eq:H0}
H_0
=&\ &
\omega_m b^\dg b 
+
\omega_0  a^\dg a^\ndg  
+
\omega_{gr} 
\sum_{i=1}^{N} 
\ketbra{r_i}{r_i}
+
\omega_{ge} 
\sum_{i=1}^{N} 
\ketbra{e_i}{e_i} 
+
\omega_p 
a^\dg_p a^\ndg_p.
\end{eqnarray} 
with the membrane phonon annihilation operator $b$ (frequency $\omega_m$),
the cavity photon annihilation operator $a$ (frequency $\omega_0$),
the Rydberg state $\ket{r_i}$ (transition frequency $\omega_{gr}$) and
the intermediate state $\ket{e_i}$ (transition frequency $\omega_{ge}$), where 
the ground state energy is set to zero.
In addition we have introduced an auxiliary cavity mode $a_p$ (frequency $\omega_p$), 
which is needed to enhance the optomechanical coupling 
without driving the atomic ensemble, as discussed below.
The interaction between the moving membrane and the atomic ensemble of Rydberg atoms is mediated by the cavity and the laser field, see Fig.~\ref{fig:1}(b), such that
the full interaction Hamiltonian reads
\begin{eqnarray}
\label{eq:Hamiltonian_Interaction}
H_\textrm{ int} = H_\textrm{ m-c} + H_\textrm{at-l} + H_\textrm{d-d},
\end{eqnarray}
where $H_\textrm{ m-c}$ describes the interaction between the membrane and
the cavity, $H_\textrm{ at-l}$ includes the interaction between the atomic ensemble 
with the cavity and also an external laser field. Finally, there is the dipole-dipole interaction $H_\textrm{ d-d}$  between the atoms in the ensemble.
In the following we give details of these interaction Hamiltonians.
\subsection{Membrane-Cavity Interaction}
The membrane-cavity interaction is described by the general expression 
for the radiation pressure Hamiltonian \cite{law_pra_radiation_pressure} 
plus an additional driving field with amplitude $\mathcal E_p$:
\begin{eqnarray}
\label{eq:radiation_pressure}
H_\textrm{ m-c}
&=& g_0 \left( b^\dg + b \right) 
\left( 
a^\dg a
+
a^\dg_p a^\ndg_p
-
a^\dg a^\ndg_p
-
a^\dg_p a
\right) 
+
i\mathcal E_p e^{-i\omega_L^m t }  a^\dg_p + \hc
\end{eqnarray}
with the radiation pressure force constant $g_0$, which is typically
small. 
In order to enhance the coupling without resonantly driving the
atomic ensemble,
we propose to pump 
an auxiliary cavity mode $a_p$ with an external laser (frequency $\omega_L^m$)
and assume this cavity mode to be detuned from the
atomic transitions. 
In the limit of an intense pumping field $\mathcal E_p$ the 
auxiliary cavity mode is 
in a coherent state: $a_p \rightarrow \alpha + a_p$ such that the
radiation pressure coupling in Eq.~\eqref{eq:radiation_pressure}
can be linearized \cite{paternostro_njp_linearization}.
The linearized membrane-cavity coupling in the rotating wave approximation then reads 
\begin{eqnarray}
H^\textrm{ lin}_\textrm{ m-c} &=& 
G \ a^\dg b e^{-i\omega_L^m t} + \hc
\label{eq:phonon_photon_Hamiltonian}
\end{eqnarray}
with $G = \alpha g_0$ ($\alpha \propto \mathcal E_p$)
as the enhanced membrane-cavity coupling. 
A detailed derivation of this beam splitter interaction Hamiltonian 
is given in \ref{app:om_coupling}. 

In the following we replace $H_\textrm{m-c}$ by the linearized Hamiltonian in Eq.~\eqref{eq:phonon_photon_Hamiltonian}, such that the interaction Hamiltonian reads: $H_\textrm{ int} = H^\textrm{ lin}_\textrm{ m-c} + H_\textrm{at-l} + H_\textrm{d-d}$.
\subsection{Atom - Light Interaction}
Benefitting from the collective enhancement of the cavity-superatom coupling is essential for the proposal.
To maximize this effect we choose to couple the transition between  
ground $\ket{g_i}$ and intermediate excited state $\ket{e_i}$ to the cavity 
and drive the transition from the intermediate excited
to the Rydberg state $\ket{r_i}$ with an external laser, cf. Fig.~\ref{fig:1}(a).
We choose this particular setup, since the external laser field can be adjusted
in its intensity  to compensate for the decrease in the dipole strength ($\sim \nu^{-3/2}$) for 
increasing effective quantum numbers $\nu$ of the Rydberg state
\cite{gallagher_review_rydberg}.\footnote{$\nu := n - \delta(n)$ includes the principal quantum number $n$ and the quantum defect $\delta(n)$.}
The atom-light interaction Hamiltonian
in dipole and rotating wave approximation reads:
\begin{eqnarray}
H_\textrm{ at-l}
=
\sum_{i=1}^{N} g_i \left(a^\ndg \ketbra{e_i}{g_i} + \hc\right)
+
\Omega 
\left( e^{-i\omega_Lt} \ketbra{r_i}{e_i} + \hc\right),
\end{eqnarray} 
where the atom-cavity coupling is denoted with $g_i$ 
and the amplitude of the external laser with $\Omega$ (frequency $\omega_L$).
Note, that $g_i$ depends on the position of the atoms. Here, we assume that the position of the atoms do not change on the timescale of the system dynamics.

\subsection{Atom - Atom Interaction}
\label{sec:at-at-int}
%
We take the dipole-dipole interaction between the highly 
excited Rydberg states into account to render an effective two-level system with the atomic ensemble:
\begin{eqnarray}
H_\textrm{ d-d} &=& 
\sum_{\substack{i,j=1\\ j>i}}^N  \Delta^{ij}_R \ \ketbra{r_i r_j}{r_i r_j} + \hc, 
\label{eq:Coulomb_Hamiltonian}
\end{eqnarray}
where $\ket{r_i r_j}$ is the doubly excited Rydberg state and $\Delta^{ij}_R$ is the induced level shift, which depends on the interatomic distance and the type
of Coulomb interaction, e.g. $\Delta^{ij}_R := -C_6/|r_i-r_j|^6$ in the Van der Waals regime \cite{pritchard_review_rydberg} with $C_6 \propto \nu^{11}$. 
An important consequence of the level shift is the Rydberg blockade mechanism, 
i.e. the level shift prevents multiple Rydberg excitations within a Rydberg blockade
radius $R_b \propto \sqrt[6]{C_6}$.
Typically, the blockade radius is in the order of microns. 
Here we assume that the Rydberg shift is large enough to allow only a 
single Rydberg excitation in the system $(\Delta^{ij}_R \gg \Omega)$, cf. Fig.~\ref{fig:1}(a).
\subsection{Dissipation Processes}
Apart from the coherent excitation transfer governed by the Hamiltonian in Eq.~\eqref{eq:Hamiltonian},
various dissipation processes contribute to the incoherent dynamics.
We include dissipation by considering the following master equation:
\begin{eqnarray}
\label{eq:full_meq}
\dot \rho &=& 
-i
\left[H,\rho\right] 
+ 
(N_m +1) 
\mathcal D [J_b]
\rho
+ 
N_m  
\mathcal D [J^\dg_b]
\rho
\\ \notag
& \ &
+ 
\mathcal D [J_a] 
\rho
+ 
\sum_{i=1}^N 
\mathcal D [J^i_e]
\rho
+ 
\mathcal D [J^i_r]
\rho .
\end{eqnarray}

The second and third contribution in Eq.~\eqref{eq:full_meq} corresponds to
the coupling of the membrane to a bath of finite temperature. 
The membrane undergoes Brownian
motion, which leads to a temperature dependent finite phonon life time.
In the Markov approximation, the phonon decay can be expressed
via the jump operator $J_b := \sqrt{\gamma_m}\  b$ with  
decay rate $\gamma_m$ and thermal occupation
of the mechanical mode $N_m \approx k_B T/\omega_m$ \cite{paternostro_njp_linearization}.
The fourth contribution denotes the cavity decay with jump operator $J_a := \sqrt{\kappa}\  a$ and cavity photon decay rate $\kappa$.
Finally, the excited states in the atomic ensemble decay radiatively.
This spontaneous emission process is described by $J^i_e := \sqrt{\Gamma_e}  \ketbra{g_i}{e_i}$ for the decay of the intermediate excited state with rate $\Gamma_e$ and by $J^i_r := \sqrt{\Gamma_r} \ \ \ketbra{g_i}{r_i}$ for the decay of the Rydberg state \cite{pedersen-pra-decay}.\footnote{We assume identical radiative decay constants for the individual atoms. Collective enhancement factors of the radiative decay constants do not change the order of magnitude in typical Rydberg ensembles, cf. \cite{pedersen-pra-decay,saffmann_review_rydberg}.} The cascaded Rydberg decay to the ground state is modelled as an effective single decay rate \cite{glaetzle_rydberg}. Furthermore, our model ignores black body radiation and super radiance effects \cite{lee_Rydberg, buechler_review}.
\newline

Until now we considered an ensemble of $N$ three-level atoms. 
In the following section we want to proceed by transforming from the microscopic description of single atoms to the macroscopic description of the Rydberg superatom. 
Therefore, we first introduce a collective basis and subsequently eliminate the intermediate state as well as the cavity degrees of freedom.
As a result, we find that the effective dynamics can be described by a Jaynes-Cummings
type of interaction.

\section{Superatom Picture: Collective Dynamics} \label{sec:collective_dynamics}

In this section we want to write the previous microscopic description of our atomic ensemble and introduce a collective description. 
In doing so we first describe the dynamics on the atomic side and formulate conditions and limits in which we obtain an effective two-level description of the ensemble - the superatom. 
In particular we discuss dissipation processes within the atomic ensemble, which lead to population of undesired non-symmetric collective states.
We find that in the regime identified below only the coherent dynamics ($g_i \rightarrow \sqrt{N} g_i$) benefits from the collective enhancement  in contrast to the radiative decay.

As a second step, we formulate an effective superatom-membrane interaction. In doing so, we eliminate the intermediate excited state of the atoms as well as the cavity degree of freedom 
 leading to an effective Jaynes-Cummings type of interaction between the Rydberg superatom and the membrane as given in Eq.~\eqref{eq:jcm_meq} in the overview.
In the last part of this section we discuss the strong coupling conditions and show
that the strong coupling regime can be reached within experimentally accessible
parameters. 

\subsection{Collective Basis}
\label{sec:collbas}

\begin{figure}[t!]
 \centering
 \includegraphics[width=0.9\textwidth]{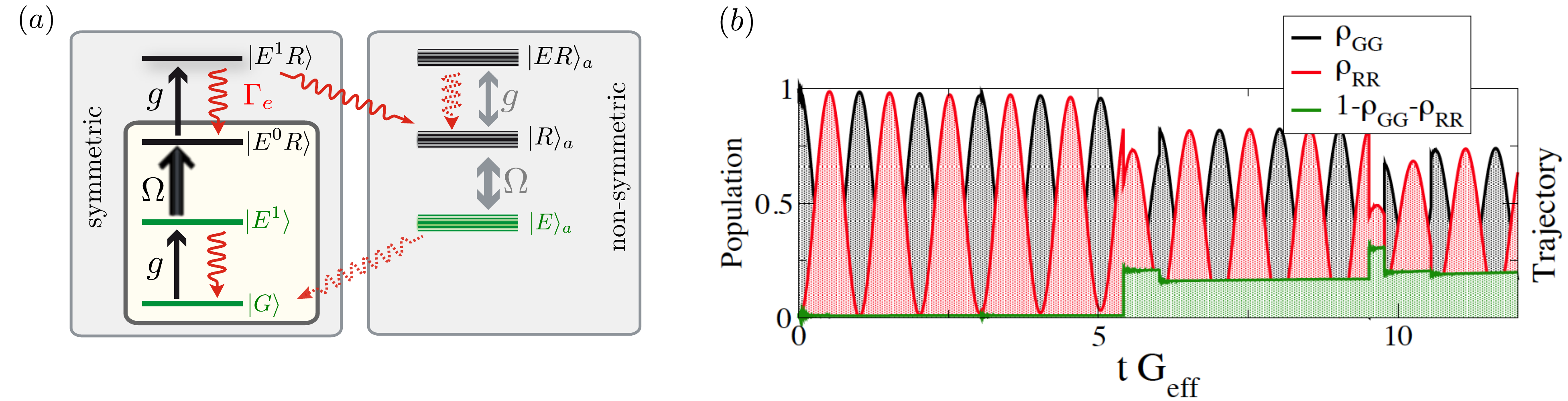}
 \caption{(a) The radiative decay of the intermediate state as an example for a mechanism which transfers population from the symmetric subspace to the non-symmetric subspace. The radiative decay from non-symmetric (symmetric) states is denoted with a dotted (solid) arrow. (b) A quantum Monte Carlo trajectory for $N=9$ three-level atoms is depicted. Clearly, jumps due to the radiative decay from symmetric states (red, black) lead to an increase of excitations in the non-symmetric subspace (green) for $tG_{\rm eff}>5$. We plot $\rho_{GG}:=\bra{G}\rho \ket{G}$ and $\rho_{RR}:=\bra{E^0R}\rho \ket{E^0R}$.}
 \label{fig:dissipative_dynamics}
 \end{figure}

In the following we introduce the collective basis by first discussing the coherent part of the dynamics, 
see Fig.~\ref{fig:dissipative_dynamics}(a).
In a second step, we extend the discussion to incoherent population
transfer and show how dissipative processes lead to population of non-symmetric states,
cf. Fig.~\ref{fig:dissipative_dynamics}(a).
\subsubsection{Coherent Dynamics}
\label{sec:cohdynSA}
The coherent excitation dynamics are governed by the Hamiltonian in Eq.~\eqref{eq:Hamiltonian}.
We assume that all atoms are initially in the ground state $\ket{G} := \ket{g_1 ... g_N}$.
The cavity photons then excite 
the atomic ensemble with coupling strength $g$
\footnote
{
We assume a setup, where the atoms couple equally to the cavity mode, i.e. $g_i = g$ 
Experimentally, this can be achieved by positioning the atoms inside 
the cavity via state-of-the-art trapping techniques as in Ref.~\cite{purdy_prl_tunable_cavity_om}.
.}
to a symmetric superposition of intermediate excited states 
\begin{eqnarray}
\ket{E^j} 
&:=& 
\frac{1}{\sqrt{N_E^j}}
\left(\sum_{i=1}^N \ketbra{e_i}{g_i} \right)^j  \ket{G}
\label{eq:coll_intermed_excited} 
\end{eqnarray}
with normalization $N^j_E := N!j!/(N-j)!$ \cite{lukin_prl_rydberg}.
From this intermediate excited states the atomic ensemble is driven by the external
laser with Rabi frequency $\Omega$ to the collective Rydberg state defined by
\begin{eqnarray}
\ket{E^j R} 
&:=&  
\frac{1}{\sqrt{N_R^j}}
 \left(\sum_{i=1}^N \ketbra{e_i}{g_i} \right)^{j}
 \left(\sum_{i=1}^N \ketbra{r_i}{g_i}\right)  \ket{G}
 \label{eq:collRydberg}
\end{eqnarray}
with normalization $N^j_R := N \cdot \ N!j!/(N-j)!$ \cite{lukin_prl_rydberg}.
Here, $\ket{E^j R}$ are the symmetric superpositions of all collective states with one atom being excited to the Rydberg state, while the other atoms are either in the ground or intermediate excited state.
Since we assume a blockade radius larger than the size of the atomic ensemble, only a single Rydberg excitation can exist within the ensemble, see Sec.~\ref{sec:at-at-int}.
%

%
In case of purely coherent dynamics only the symmetric states, Eqs.~\eqref{eq:coll_intermed_excited}-\eqref{eq:collRydberg}, are populated
by the interaction Hamiltonian.\footnote{Note, this is only valid as long as all atoms couple equally to the cavity mode: $g_i = g$.}
However, due to dissipation processes non-symmetric states also become populated, as discussed in the following.

%
\subsubsection{Incoherent Dynamics}
%
In Fig.~\ref{fig:dissipative_dynamics}(a), 
we illustrate how dissipative processes transfer population from the symmetric 
 (left side) to the non-symmetric subspace (right side). 
The non-symmetric subspace consists of all states that are not permutation invariant, such as   
$\ket{E^1}_a := \left( \ket{e_1  g_2} - \ket{g_1 e_2} \right)/\sqrt{2}$ (in the case of two atoms).

To give an instructive example how these dissipation processes lead to population of non-symmetric states, we consider the radiative decay from 
the intermediate excited state $\Gamma_e$ in the case an ensemble of two atoms.
Non-symmetric states are populated by this process, 
since the radiative decay acts on the individual atoms 
and not collectively on the whole ensemble.
Beginning with both atoms in the ground state $\ket{G}$, the atom-cavity coupling creates a superposition state $\ket{E^1}= \left( \ket{g_1  e_2} + \ket{e_1 g_2} \right)/\sqrt{2}$ with one atom excited to the intermediate state, see Sec.~\ref{sec:cohdynSA}.
This superposition state can either decay back to the ground state by spontaneously emitting 
a photon (with probability proportional to the radiative decay constant $\Gamma_e$),
or it is driven by the external laser to a superposition state with a 
single Rydberg excitation $\ket{E^0R}= \left( \ket{g_1  r_2} + \ket{r_1 g_2} \right)/\sqrt{2} $, cf. Fig.~\ref{fig:dissipative_dynamics}(a).
Then, if the atomic ensemble absorbs another cavity photon, 
a doubly excited state is created with each an excitation in the intermediate
and Rydberg state: $\ket{E^1R}= \left( \ket{r_1  e_2} + \ket{e_1 r_2} \right)/\sqrt{2} $.
This state either couples via the cavity interaction again to the symmetric 
collective state $\ket{E^0R}$ or decays radiatively. 
Considering the latter case, we find that the radiative decay of the intermediate excited state of e.g. the first atom $J_e^1=\sqrt{\Gamma_e}\ketbra{g_1}{e_1}$, leads to a single Rydberg excitation: $ J_e^1 \ket{E^1R}\propto \ket{g_1 r_{2}}/\sqrt{2}$. 
Rewriting this state in the collective basis yields:
 $\ket{g_1 r_{2}} = \left( \ket{E^0R} - \ket{R}_a \right)/\sqrt{2}$, which corresponds to a superposition of a symmetric and a non-symmetric state, see Fig~\ref{fig:dissipative_dynamics}(a).
The latter is defined by $\ket{R}_a := \left( \ket{r_1  g_2} - \ket{g_1 r_2} \right)/\sqrt{2}$.
As an illustration, we give in Fig.~\ref{fig:dissipative_dynamics}(b) a typical quantum Monte Carlo trajectory \cite{gardiner-book} computed
for $N=9$ three-level atoms.
Clearly, the population transfer from the symmetric subspace (red, black) 
to the non-symmetric subspace (green) is associated with a 
quantum jump due to the radiative decay, e.g. at $t G_{\rm eff} \approx 5$.
The source of this population transfer is the strong radiative decay of the intermediate excited state.
In consequence, if such a population of the intermediate excited state is suppressed,
the restriction to the symmetric collective basis is a good approximation. The detailed limits in which the intermediate state can be eliminated are discussed in the following section, see \ref{app:radiative_decay_suppression} for an instructive example how to suppress the radiative decay via detuned excitation, and \ref{app:meq_symmetric_basis} for the full master equation in the symmetric collective basis.

\subsection{Superatom-Membrane Interaction} \label{sec:adiabatic_elimination}

%
In this section we derive a Jaynes-Cummings type of interaction between
the Rydberg superatom and the moving membrane as given in the overview, cf. Eq.~\eqref{eq:jcm_meq}.

%
\begin{figure}[t!]
\centering
\includegraphics[width=0.75\textwidth]{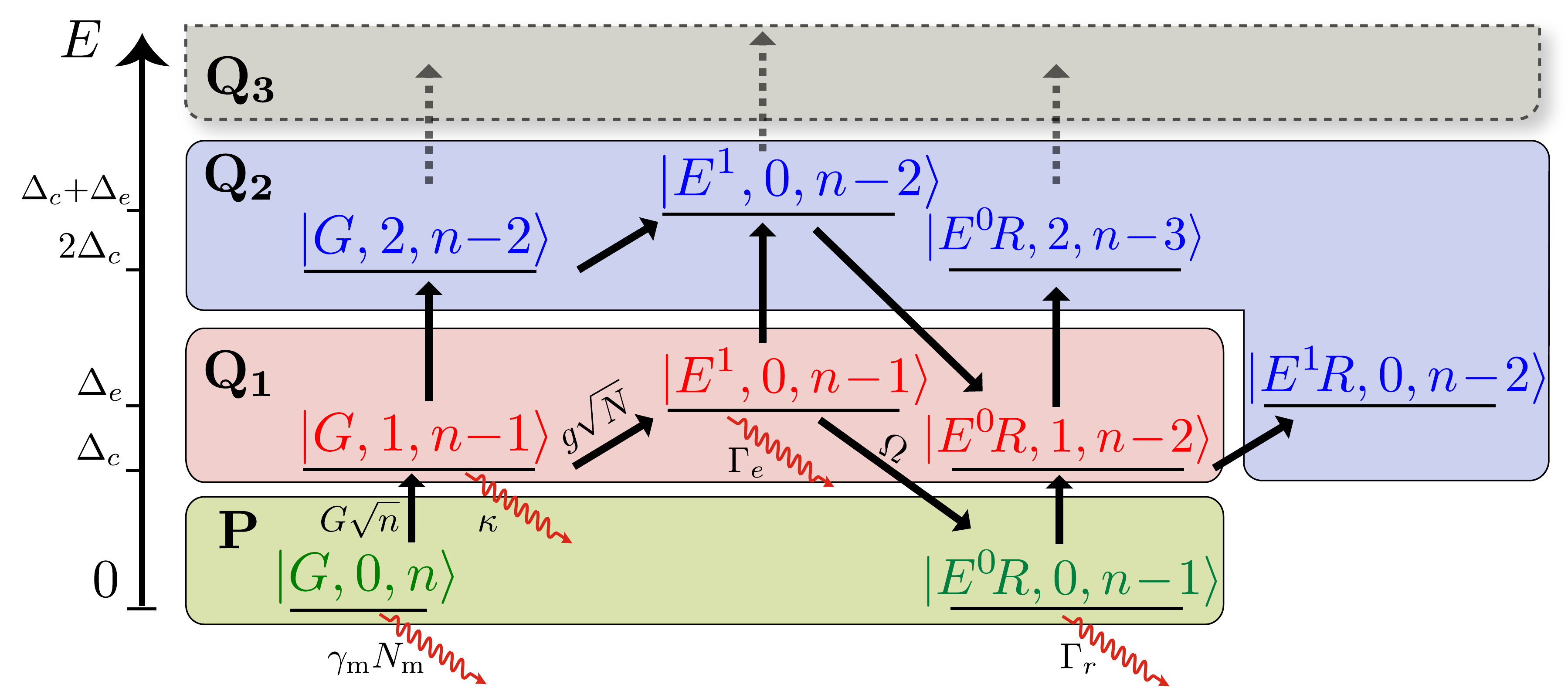}
\caption{Multiple excitations lead to a complex coupling between the excitation manifolds of the atomic ensemble. We divide the Hilbert space in the relevant part (green) and the irrelevant part (blue, red, grey). The irrelevant part is also divided into different parts, depending on how many times the Hamlitonian needs to be applied on the wave function. Starting in the relevant part of the Hilbert space: $Q_1$ (red) consists of states, which are reached for a single application of $H$, $Q_2$ (blue), if the Hamiltonian needs to be applied two times, etc.}
\label{fig:scheme_multi_ex}
\end{figure}

%
To suppress  the cavity loss and the radiative decay from the intermediate state, which 
 justifies the restriction to the symmetric
basis, we assume that all states with a cavity or with an intermediate state
excitation to be detuned 
with $\Delta_c = \omega_0- \omega_L^m - \omega_m $ and $\Delta_e = \omega_e - \omega_L^m - \omega_m$, respectively, by choosing a resonance condition $\omega_r = \omega_L + \omega_L^m + \omega_m$.

For large detunings we can treat the dynamics in the adiabatic limit given by
\begin{eqnarray}
\Delta_c,\Delta_e \gg G \sqrt{n}, \Omega, \Gamma_e, \kappa , \qquad
\Delta_c \Delta_e \gg g^2 N n ,
\label{eq:adiabatic_condition}
\end{eqnarray}
where $n$ is the number of excitations in the system.
In the spirit of perturbation theory we derive the effective dynamics of the subspace consisting of the states: $\ket{G,0,n}$ and $\ket{E^0R,0,n-1}$, cf. Fig.~\ref{fig:scheme_multi_ex}(green). 
Here, we have introduced the notation $\ket{A,i,j}$, where the atomic ensemble 
is in state $A$ with $i$ photons and $j$ phonons. 
By using the projection operator method \cite{gardiner-book} we derive an effective master equation for the $n$ excitations subspace:
\begin{eqnarray}
\notag
\partial_t \rho^n_s
& = &
-i\left\lbrack H^n_s,\rho^n_s \right\rbrack 
+ 
( \Gamma_r + \Gamma_r^{\rm eff} )
\mathcal D\left[ \ketbra{G,n-1}{R,n-1} \right] \rho^n_s 
\\
& \ &
+
\left(  ((N_m +1) \gamma_m + \gamma_m^{\rm eff} ) 
\mathcal D [ b ]
+
N_m  \gamma_m
\mathcal D [b^\dg]
\right) \rho^n_s,
\label{eq:n_eff_MEQ}
\end{eqnarray}
where we have defined $\ket{R,n} := \ket{E^0R,0,n}$ and $\ket{G,n} := \ket{G,0,n}$.
The radiative decay $\Gamma_e$ of the 
intermediate state and the cavity loss $\kappa$ act as an effective phonon decay rate
$\gamma_m^\text{\rm eff} \approx \kappa (G/\Delta_c)^2 $ 
and as an effective decay of the Rydberg 
state $\Gamma^\text{eff}_r \approx  \Gamma_e (\Omega/\Delta_e)^2$, respectively. 
In \ref{app:effective_model} we discuss in more detail the derivation and
give the analytical expressions for the single-excitation limit ($n=1$).

The effective Hamiltonian then reads
\begin{eqnarray}
H_s^n 
 = &
-&n \ \Delta_G \ketbra{G,n}{G,n}
-\Delta_\Omega  \ketbra{R,n-1}{R,n-1} 
\notag \\
&+& 
\sqrt{n}
G_\text{eff} 
\left(\ketbra{R,n-1}{G,n} 
+ 
\hc\right)
\label{eq:n_eff_Hamiltonian}
\end{eqnarray}
with the effective coupling and dispersive shifts defined by
\begin{eqnarray}
\label{eq:Geff}
G_{\rm eff} 
&\approx&
\sqrt{N} \frac{g G \Omega}{\Delta_{e}\Delta_{c}}
,
\quad
\Delta_G \approx \frac{G^2}{\Delta_c}
,
\quad
\Delta_\Omega \approx \frac{\Omega^2}{\Delta_e}.
\end{eqnarray}
Clearly, with this Hamiltonian
our goal to engineer a Jaynes-Cummings type of interaction between the
membrane and the Rydberg superatom is achieved, cf. Eq.~\eqref{eq:jcm}.
In Eq.~\eqref{eq:Geff} we see a superatom-membrane coupling that scales with the square root of the number of atoms. 
The detuned setup allows the suppression of the dissipative processes, which in consequence are not collectively enhanced. However, this comes at a price of a slower excitation transfer rate $G_{\rm eff}$. 
Remarkably, we can
fully benefit from the superatom imposed enhancement factor.
As a consequence, for very high numbers of atoms $N$ the dissipation
processes become negligible and strong coupling is possible.
Finally, we remark, that for negligible dispersive shifts $(\Delta_G,\Delta_\Omega)$ compared
to the effective coupling $G_{\rm eff}$ we can write $H_\text{s} = \sum_n H^n_\text{s} \approx H_{\rm JCM}$
as given in the overview in the corresponding rotating frame.
However, in a regime where the dispersive shift $\Delta_G$ is large compared to the phonon heating $\gamma_m N_m$, i.e. for very low effective temperatures, the different excitation manifolds can be addressed separately. This is due to the fact that the dispersive shift scales with the number of excitations. It would we very interesting to go the such a parameter regime in the spirit of non-classical state preparation.

\subsection{Discussion: Strong Coupling Regime}\label{sec:strong_coupling}
%
The strong coupling regime is  reached if the effective coupling $G_\text{eff}$ outrivals the losses, 
i.e the effective phonon decay rate $\gamma_m^\text{eff}$, the decay rates of the
Rydberg level $\Gamma_r^\text{eff},\Gamma_r$ and the coupling of the phonons to a thermal environment
\begin{eqnarray}
\label{eq:strong_coupling}
G_\text{eff}
\gg \gamma^\text{eff}_m, \Gamma^\text{eff}_r,  \Gamma_r, \gamma_m (N_m+1).
\end{eqnarray}
By using the definition of the effective coupling and decay constants, and further assuming $G=\Omega$ and $\Delta_c = \Delta_e$, we can reformulate the condition in \eqref{eq:strong_coupling} to
\begin{eqnarray}
\sqrt{N}g
&\gg& 
\kappa
, 
\Gamma_e 
, 
\frac{\Delta^2_e }{\Omega^2}
\Gamma_r
,
\frac{\Delta^2_c }{G^2}
\gamma_m
(N_m+1).
\label{eq:strong_coupling_condition}
\end{eqnarray}
We have on the LHS the atom-cavity coupling enhanced by the square root of the number
of atoms in the ensemble.
In contrast, on the RHS the losses are not enhanced and strong coupling can be reached for
experimentally accessible parameters.
As an example, we use the parameters of state-of-the-art experiments in cavity optomechanics \cite{kippenberg_science_laser_enhancement,joeckel_apl_mechanical_dissipation} 
and Rydberg cQED \cite{brennecke_nature_bec_cqed}.
Strong coupling can be achieved for 
a high finesse cavity with a cavity loss $\kappa=2\pi\times1$MHz and
an atom-cavity coupling of  $g=2\pi \times 1$MHz, 
laser amplitude $\Omega=2\pi\times10$MHz for $^{87}$Rb atoms, using e.g. 
transitions from the $5$S$_{1/2}$ ground state to an intermediate state $5$P$_{3/2}$
with a life time of $\Gamma_e=2\pi\times3$MHz  and a transition to the Rydberg
state $60$S$_{1/2}$ with a life time $\Gamma_r \approx \Gamma_e/1000$ as in \cite{brennecke_nature_bec_cqed,saffmann_review_rydberg,pritchard_review_rydberg}, and $G=2\pi\times10$MHz 
and a phonon decay in order of magnitude of $\gamma_m (N_m+1) \approx 10$kHz as in \cite{thompson_nature_membrane,joeckel_apl_mechanical_dissipation,stannigel_pra_transducers}.
The number of Rydberg atoms in typical experiments 
\cite{schauss_nature_2D_rydberg_gas,brennecke_nature_bec_cqed} ranges up to several 
$1000$.

The fact that strong coupling can be achieved between the Rydberg superatom and the membrane
is the main result of this article.
In the next section, we discuss that our hybrid system can also be realized in
a modular long-distance setup.

\section{Long-distance Superatom-Membrane Coupling}\label{sec:laser_mediated}
%
Cavity-mediated coupling between atoms and a mechanical oscillator as described in Sec.~\ref{sec:cavity_mediated} and \ref{sec:collective_dynamics} is very demanding as it requires to combine ultra high vacuum (needed for experiments with cold atoms) with cryogenic environment (needed for experiments with micromechanical systems). 
Motivated by this we propose and discuss in this section an alternative setup that does not require a cavity mediated coupling.

The setup we have in mind is depicted in Fig.~\ref{fig:1}(c) and is similar to Hammerer 
et al. \cite{hammerer_pra_optical_lattices_om}. 
Here, we have on side the moving membrane with frequency $\omega_m$ in a cryogenic environment, on the other side we have the ensemble of Rydberg atoms, which eventually form the Rydberg superatom, in a vacuum chamber. 
Both systems can in principal be spatially separated in the order of meters. 
The atomic ensemble is driven by a laser with frequency $\omega_L$ perpendicular to the $z$-axis, and we assume that the two-photon excitation scheme to excite the Rydberg superatom is similar to the cavity-mediated case, see Fig.~\ref{fig:1}(a) inset. Further we choose the resonance condition such that a Rydberg excitation $\omega_r$ emerges from a two-photon process with one photon being a sideband photon with frequency $\omega_L^m+\omega_m$, i.e. $\omega_r = \omega_L + \omega_L^m+\omega_m$ as in Sec.~\ref{sec:cavity_mediated}.

The coupling between the moving membrane and the superatom is mediated by a laser with frequency $\omega_m^L$. As discussed in Ref.~\cite{vogell_unpublished} the coupling yields a cascaded dynamics, i.e. the order in which the systems interact with each other is relevant, which is due to the fact that the setup is only driven from one side.
First, we assume the Rydberg superatom is initially in the ground state such that an incoming coupling laser photon $\omega_L^m$ together with the a pump laser photon $\omega_L$ is not resonant with the Rydberg level exciation, and therefore does not interact with the atomic ensemble. 
The coupling laser then interacts with the moving membrane and is reflected with imprinted sidebands at $\omega_L^m \pm \omega_m$ due to the motion of the membrane at frequency $\omega_m$. 
Subsequently, the positive sideband can interact with the atomic ensemble and excite the Rydberg superatom.
In contrast, an excited Rydberg superatom can emit sideband photons at $\omega_L^m + \omega_m$ in the direction of the membrane such that it feels a change in radiation pressure.

Cavity-enhanced long-distance coupling as proposed in Ref.~\cite{vogell_pra_cavity_enhancement} has the advantage of a membrane-light coupling constant that is enhanced by the finesse. To also benefit from this enhancement  
 we extend our proposal in the following to a so-called membrane-in-the-middle configuration \cite{thompson_nature_membrane} as depicted in Fig.~\ref{fig:1}(d).

In the following we first present the full Hamiltonian for such long-distance setup. We further give the master equation of the effective Rydberg superaton-membrane coupling. The derivation of this effective master equation is outlined in \ref{app:long_distance_model}, where we crucially rely on the formalism as in Ref.~\cite{vogell_unpublished}.
As a result, we find that long-distance coupling of the superatom to the membrane is possible and features a (atomic) position-dependent
coupling, which allows to switch coupling and dissipation channels on and off. 
Finally, we find a limit in which we recover a similar master equation as the one in the cavity-mediated case, see Eq.~\eqref{eq:n_eff_MEQ}. However, the benefits of cavity-mediated long-distance coupling come at the price of an additional dissipation channel due to the membrane-light diffusion.

\subsection{Hamiltonian}
The Hamiltonian for the long-distance setup as depicted in Fig.~\ref{fig:1}(d) is given by
\begin{align}
\tilde{H}=\tilde{H}_0+\tilde{H}_{\rm m-f}+\tilde{H}_{\rm at-f},
\end{align}
where the free evolution is
\begin{align}
\label{eq:H0tilde}
\tilde{H}_0=\hbar \omega_m b^\dag b+ \hbar \int\! \text{d}\omega \, \omega \,  c_\omega^\dg c_\omega^\ndg  
+
\omega_{gr} 
\sum_{i=1}^{N} 
\ketbra{r_i}{r_i}
+
\omega_{ge} 
\sum_{i=1}^{N} 
\ketbra{e_i}{e_i}.
\end{align}
The interaction Hamiltonian consists of the membrane-light field $\tilde{H}_{\rm m-f}$ and the superatom-light field interaction $\tilde{H}_{\rm at-f}$. The former is given by Eq.~(13) in Ref.~\cite{vogell_pra_cavity_enhancement} with a light field quantized similar as presented there. The superatom-light field interaction includes both, the coupling of the classical pump laser $\omega_L$ with Rabi frequency $\Omega_L$ to the transition from intermediate excited to Rydberg state as well as the interaction of the coupling laser $\omega_L^m$ with Rabi frequency $\Omega_L^m$  to the transition from ground to intermediate excited state.t
Note, compared to the cavity-mediated case, where the membrane coupled to a single cavity mode, we have in the long-distance case a full continuum of field modes centered around the coupling laser frequency $\omega_L^m$.

In \ref{app:long_distance_model} we give the expression for the full Hamiltonian and further outline of the derivation, which is similar to the methods used in \cite{vogell_unpublished}, to obtain an effective master equation for the superatom-membrane system. In the following section we present the resulting master equation and discuss a limit, where we recover a similar master equation as in the cavity-mediated case in Eq.~\eqref{eq:n_eff_MEQ}.

\subsection{Master Equation Dynamics}

Concluding the derivation in \ref{app:long_distance_model} we find that the Hamiltonian in the collective basis, cf. Sec.~\ref{sec:collbas}, reads
\begin{eqnarray}
\label{eq:Hcollind}
H^{\rm coll}_{\rm ind}
&=&
-\hbar \Delta_\text{at}
 \ \sin(2k_L^m \bar{z}_j) \ \sigma_{R R} 
+
\hbar 
\bar{G}_{\rm eff} \left( \cos(k_L^m \bar{z}_j ) \ b^\dg  \sigma_{G R} + \hc \right)
\end{eqnarray}
with dispersive shift $\Delta_\text{at}$, wave vector $k_L^m$,
the mean position of the atoms $\bar{z}_j$, $\sigma_{ab}=\ketbra{a}{b}$ and effective long-distance coupling constant $\bar{G}_{\rm eff}$.
The corresponding master equation is given by 
\begin{eqnarray}
\label{eq:HcollindMEQ}
\dot \rho
&=&
-\frac{i}{\hbar} \left[ H^{\rm coll}_{\rm ind} ,\rho\right] + \frac{\gamma_m^\text{diff}}{4}  \mathcal{D}\left[b^\dg \right]  \rho \\
& \ & 
+2 \Delta_\text{at} \sin^2(k_L^m \bar{z}_j) 
\left( 
2 N \ \sigma_{GR} \rho \sigma_{RG} -  \sigma_{RR} \rho - \rho \sigma_{RR} 
\right) \notag\\ \notag
& \ & 
+\frac{\bar{G}_{\rm eff}}{\sqrt{N}} \sin(k_L^m \bar{z}_j) 
\left( 
2 b \rho \sigma_{RG}   
- \sigma_{RG} b^\ndg \rho -  \rho \sigma_{RG} b^\ndg
  + \hc \right),
\end{eqnarray}
where $\gamma_m^\text{diff}$ is the membrane-light diffusion as defined in Ref.~\cite{vogell_pra_cavity_enhancement}. 
Note that all dissipation channels concerning the atoms as well as the effective coupling in Eq.~\eqref{eq:Hcollind} are position dependent. 
This due to coupling to internal states of the atomic ensemble in contrast to coupling to the motional states of the atoms, where a Lamb-Dicke expansion is applied \cite{vogell_pra_cavity_enhancement}.
Therefore,
by appropriately choosing the position of atoms, dissipation channels
can be switched.

As a specific example we consider the case, where the coupling in the Hamiltonian in Eq.~\eqref{eq:Hcollind} gets maximum, i.e. $k_L^m z \ll 1$, and thereby recover a similar master equation as in the cavity mediated case, see Eq.~\eqref{eq:n_eff_MEQ}. 
Further, taking dissipation due to the radiative decay
of the Rydberg level and the heating of the membrane into account we obtain\begin{eqnarray}
\notag
\dot \rho
&=&
-i
\bar{G}_{\rm eff}
\left[ 
 b^\dg  \sigma_{G R}
+
\sigma_{RG} b ,\rho\right] 
+ 
\frac{\gamma_m^\text{diff}}{4} 
\mathcal{D}\left[b \right] \rho \notag \\
&\ & 
+
\frac{\gamma_m}{2} \left(N_m+1 \right) 
\mathcal{D}\left[b \right]  \rho 
+
\frac{\gamma_m}{2} N_m 
\mathcal{D}\left[b^\dg \right]  \rho
+ 
\frac{\Gamma_r}{2} \mathcal{D}\left[ \sigma_{GR} \right] \rho.
\label{eq:meq_long_distance_sine_sin}
\end{eqnarray}
In contrast to the cavity-mediated case we obtain an additional dissipation channel on the side of the membrane with rate $\gamma_m^\text{diff}$ resulting from the elimination of the coupling field, cf. \cite{vogell_pra_cavity_enhancement}.

Summarizing the long-distance version of the proposed setup we find that in a certain limit we can recover the previous results, cf. \ref{sec:collective_dynamics}. However, other limits can also be engineered due to the position dependency of the dissipation and the coupling.

\section{Toolbox for Cavity Optomechanics}\label{sec:toolbox}
%
In this section, we point out how a 
Rydberg superatom can be used as a toolbox for cavity optomechanical experiments.
Our hybrid system allows engineered 
dissipation of the Rydberg superatom and
can be utilized to cool the membrane or to read out its
state via spectroscopy techniques.
As concluded in Sec.~\ref{sec:collective_dynamics} the proposed setup is not restricted to a Jaynes-Cummings type of
interaction. 
For example, by increasing (decreasing) the intensity of the 
external laser $\Omega$, one can decrease (increase) the
Rydberg blockade radius and thus change the number of superatoms
in the cavity.
Thereby, a superatom based Tavis-Cummings model \cite{shore_review_jaynes_cummings} 
could be realized.
In such a setup, superatoms can become entangled with each other,
even when they do not interact directly, and multipartite entanglement is
generated \cite{tavis_cummings_pra_entanglement}. 
Vice versa, by including a second mechanical mode in the system, a multi-mode
Jaynes-Cummings model is realized.
This can be done by either addressing two modes on a single membrane
or by inserting a second membrane in the system.
Analogously to the multi-atom case, the multi-mode Jaynes-Cummings
model allows the preparation of shared entanglement \cite{PhysRevA.64.050301,multi_mode_jcm_entanglement}.
Below, we give three examples for possible applications: Fock state transfer,
Superatom-mediated cooling and non-classical state preparation
for a membrane initially in a thermal state.

\subsection{State transfer} 
%
In the strong coupling limit, a transfer of a Fock state with $n=1$  
from the membrane to the Rydberg superatom and vice versa
can be achieved with a very high fidelity.
For example,  
to swap the excitation from the Rydberg superatom
to the membrane $\ket{R,0} \rightarrow \ket{G,1}$ a time $t_g = \frac{\pi}{2G_\text{eff}}$ is necessary and the fidelity $\mathcal F = \bracket{G,1}{\Psi_f(t_g)}$ scales as
\begin{eqnarray}
\mathcal F 
\approx 1 
-
\frac{\pi}{2 G_\text{eff} }
\left(
4N_m\gamma_m+\gamma_m 
+
\Gamma^\text{eff}_r+\Gamma_r
\right).
\label{eq:fidelity_single_excitation}   
\end{eqnarray}
Due to the strong coupling condition in Eq.~\eqref{eq:strong_coupling_condition},
it is clear that a high atom number leads to a coupling element that outrivals the losses in the system, and thus 
high fidelities are possible.
However, to obtain high fidelities it is necessary to have a membrane cooled to the ground state in order 
to avoid degrading effects due to heating.

\subsection{Sympathetic Cooling} 
%
Considerable
experimental work has been drawn to realize ground state cooling of the
mechanical system \cite{o2010quantum,teufel2011sideband,chan2011laser}.
In addition to the standard optomechanical cooling via the cavity decay \cite{wilson2008cavity}
we discuss here sympathetic cooling \cite{vogell_pra_cavity_enhancement} 
of the membrane by utilizing the well-developed AMO toolbox to 
extract Rydberg excitations from the atomic ensemble \cite{saffman_loading_extracting}.

The sympathetic cooling of the membrane is achieved by the following steps:
first an excitation is transferred from the membrane via the coherent
Jaynes-Cummings coupling to the Rydberg superatom.
In a second step, a $\pi$ pulse with amplitude $\Omega_d$ in resonance with an auxiliary ground state $\ket{g^\prime}$ of the atoms
is applied and deexcites the atom to this ground state:
$H_{\rm cool} = \Omega_d \left( \ketbra{R}{g^\prime} + \text{H.c.}  \right)$.
The superatom is then its ground state with one atom less $N \rightarrow N-1$,
and the atom from in the auxiliary ground state can be removed according to the 
jump operator: $J_\text{\rm cool} := \sqrt{\gamma_{cl}} \ketbra{\text{vac}}{g^\prime} $.
Eliminating the auxiliary ground state with $\gamma_{cl}\gg \Omega_d$
and $N\gg 1$, we can write an effective dissipation: 
\begin{eqnarray}
\partial_t \rho = 
-i\left[ H_{\rm s}^n,\rho \right] 
+ 
\mathcal{D}\left[ \sqrt{\gamma^R_\text{cool}}  \ketbra{\text{vac}}{R} \right]\rho
\end{eqnarray}
with $\gamma^R_\text{cool} := \Omega_d^2/\gamma_{cl}$ and $H_{\rm s}^n$ in Eq.~\eqref{eq:n_eff_Hamiltonian}.
The dissipation is steered by the laser with Rabi frequency $\Omega_d$ 
and can be switched on and off to remove excitations from the system.
In a third step, another excitation from the membrane is transferred and  
the cooling cycle can be repeated.
Using this one could gradually cool the membrane and 
thereby prepare it its ground state.
In the strong coupling limit $G_{\rm eff} \gg N_m \gamma_m$,
the steady state phonon number $n_{\rm s}$ scales with  \cite{wallquist_pra}
\begin{eqnarray}
n_{\rm s}
\approx 
2 N_m \gamma_m 
\left(
\frac{1}{G_{\rm eff}}
+
\frac{1}{\gamma^R_{\rm cool}}
\right) .
\end{eqnarray}
Providing a strong coupling, the thermal occupation of the
membrane  can be reduced to very low mean phonon number
and in principle to the ground state.
%

\subsection{State preparation} 
\begin{figure}[t!]
\centering
\includegraphics[width=0.85\textwidth]{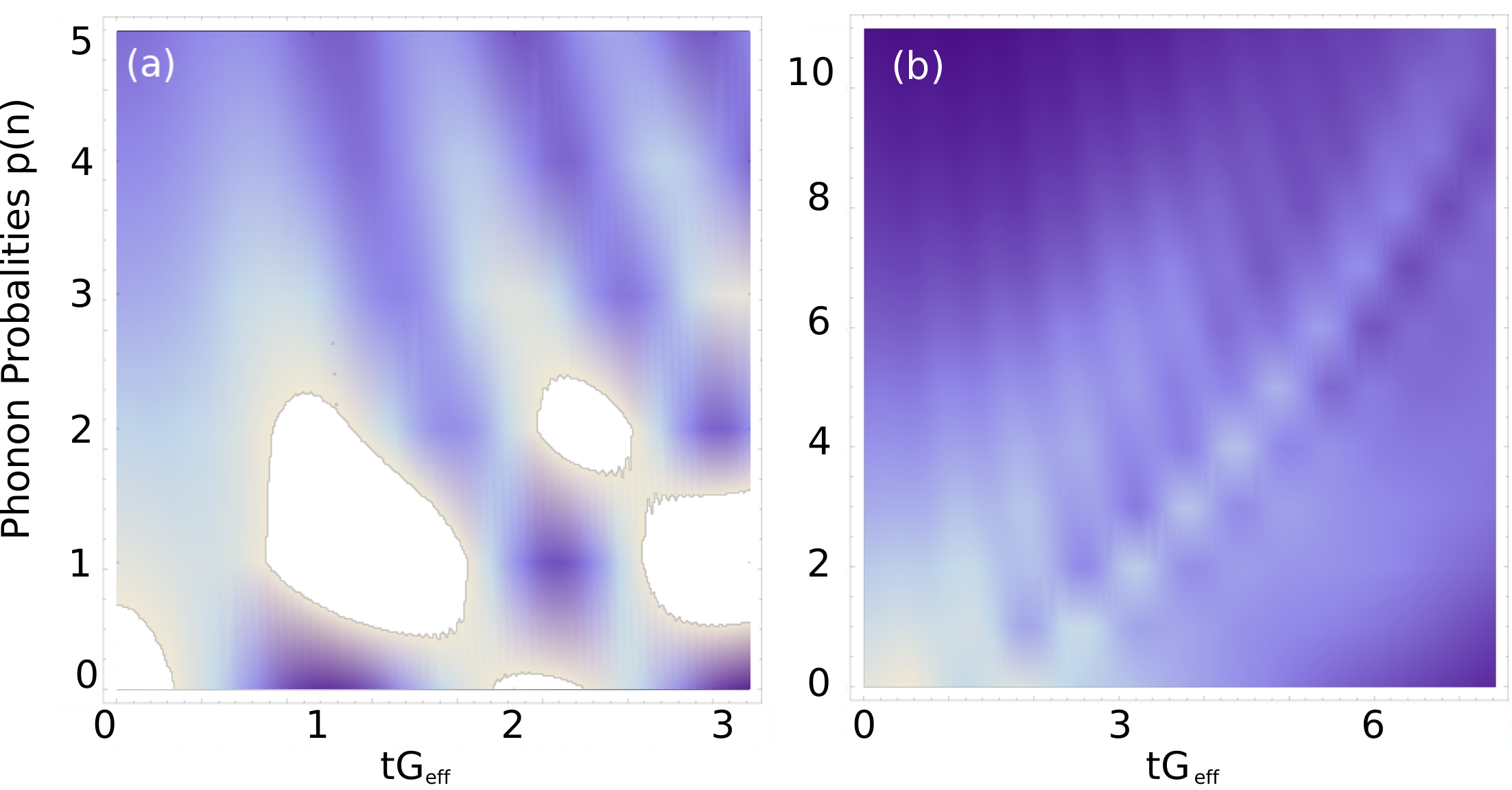}
\caption{Time evolution of the phonon probability distribution of a membrane for a heat bath: ( a ) $N_m=1$ and ( b ) $N_m=7$. Due to the strong coupling to a coherently pumped Rydberg superatom, the initial Bose-Einstein distribution is changed into a non-classical state with increasing Fock phonon numbers. If the thermal occupation is high, a stronger pumping is necessary to generate a non-classical state due to the strong dissipative dynamics in between the phonon manifolds.}
\label{fig:non_classical_state}
\end{figure}

%
The possibility to coherently drive the superatom and to switch 
the interaction between the membrane and the superatom allows
a great control of the membrane - superatom interaction along the line of
nonlinear quantum optomechanics via intrinsic two-level defects \cite{ramos_prl_defects}.
This gives rise to the prospect of deterministically preparing mechanical states by suitable protocols.
In Fig.~\ref{fig:non_classical_state}, we numerically evaluate the
master equation \eqref{eq:meq_long_distance_sine_sin} of the long 
distance coupling for a membrane coupled to a heat bath with
mean phonon occupation number of $N_m = 15$.
At $t=0$, a laser is switched on and drives the atomic
ensemble, continuously creating a Rydberg excitation.
As a result of the Jaynes-Cummings type of interaction, the Rydberg excitation is transferred to the membrane
and the phonon distribution changes from a Bose-Einstein 
to a non-classical distribution, centered around the mean value
of $p_n=1$ as a signature of thermal phonon distribution
doped with one Fock phonon.
With increasing time, Rabi oscillations become visible.
First, the main contributions arise between the
states $\ket{G,1}$ and $\ket{R,0}$ for $(G_{\rm eff}t \approx 5)$.
Due to the pumping process, the mean phonon number
increases and the main contributions is changed to
the Rabi oscillation between the states $\ket{G,2}$ 
and $\ket{R,1}$ for $(G_{\rm eff}t \approx 10)$.
These dynamics show, that even for a membrane,
which is not in the ground state, the strong coupling
allows a non-classical state preparation.

%

\section{Conclusion} \label{sec:conclusion}
We investigated a system consisting of a membrane and a Rydberg superatom formed by an atomic ensemble.
Due to the superatom imposed collective enhancement factor, we show
that strong coupling between the superatom and the membrane can be achieved
for parameters within the range of current experiments.
The strong coupling regime can be reached cavity-mediated or in a long
distance setup and allows to utilize the Rydberg superatom as a
toolbox for optomechanical experiments such as ground state cooling
and (non-classical) state preparation.
This hybrid system constitutes a new feasible implementation for a qubit strongly coupled to a harmonic oscillator. 
\\ 
After completing this work we became aware of  \cite{bariani2013single}, where also
nanomechanics is coupled to a Rydberg ensemble. In contrast to that work, we focus here within a fully microscopical model for a single Rydberg state
on the competition between the incoherent and coherent part of the dynamics.
\ \\
This project was supported by the ERC Synergy Grant UQUAM, the SFB FoQuS (FWF Project No. F4006-N16) and the Marie Curie Initial Training Network COHERENCE.
A.C. acknowledges gratefully support from Alexander-von-Humboldt foundation through the Feodor-Lynen program.

\begin{appendix}
\section{Enhancement of the Optomechanical Coupling}\label{app:om_coupling}
In this section we demonstrate one possible way to obtain the enhancement of the optomechanical coupling without also driving the atomic transitions. 

We investigate a two mode cavity system with frequencies $\omega_0$ and $\omega_p$, which mediates an excitation transfer between a moving membrane and an atomic ensemble.
To enhance the radiation pressure coupling $g_0$ between the membrane with frequency $\omega_m$ and the cavity modes \cite{law_pra_radiation_pressure}, a laser with frequency $\omega_L^m$ and Rabi frequency $\mathcal E_p$ pumps the cavity from the right, see Fig.~\ref{fig:1}(a).  
A two-mode cavity set-up is assumed to avoid a strongly driven transition between the ground and intermediate excited state of the atoms.
In a corresponding rotating frame the Hamiltonian of the full system reads $(\hbar=1)$ :
\begin{eqnarray}
H
=&\ &
\Delta_p a^\dg_p a^\ndg_p 
+ 
\Delta_c  a^\dg a^\ndg  
+
\omega_m b^\dg b 
+
\Delta_e 
\sum_{i=1}^{N} 
\ketbra{e_i}{e_i}  \\ \notag
&+&
g  \sum_{i=1}^{N} \left(a^\dg \ketbra{g_i}{e_i} + \hc\right)
+
\Omega
\left( \ketbra{e_i}{r_i} + \hc\right)
 \\ \notag
&+&
\sum_{\substack{i,j=1\\ j>i}}^N \Delta_R^{ij} \ \ketbra{r_i r_j}{r_i r_j} + \hc 
 \\ \notag
&+&
g_0 (b^{\dg} + b) 
\left( 
a^\dg a 
+  
a^\dg_p a^{\ndg}_p 
-
a^\dg a^{\ndg}_p \ e^{i\omega_m t}
-
a^\dg_p a \ e^{-i\omega_m t} 
\right)    \\ \notag
&+&
i\mathcal E_p \  a^\dg_p 
+ 
\hc ,
\end{eqnarray} 
with the detuning of the pump cavity field $\Delta_p = \omega_p - \omega_L^m$,  the detuning of the cavity field that couples to the atomic ensemble $\Delta_c = \omega_0 - \omega_L^m - \omega_m$, the detuning of the intermediate excited state $\Delta_e = \omega_e - \omega_m - \omega_L^m$, and the resonance condition $\omega_r=\omega_m +\omega_L^m+ \omega_L $. 
Since the pump cavity field $a_p$ is far detuned from the transition between ground and intermediate excited state of the atoms, the coupling between them is neglected.
\\ 
The Langevin formalism accounts for the complete open dynamics of the system \cite{gardiner-book}.
Explicitly, the quantum Langevin equations (QLEs) of the cavity modes $(a^\dg_p,a^\dg)$ and the membrane $(b^\dg)$ read \cite{paternostro_njp_linearization}:
\begin{eqnarray}
\partial_t a_p
&=&
-(i\Delta_p + \kappa) a_p
-
i
g_0 a_p (b^\dg + b) \\ \notag
& \ &
+i 
g_0 a (b^\dg + b) \  e^{-i \omega_m t}
+\sqrt{2\kappa} a_{\text{in}} \\
\partial_t a
&=&
-\left(i \Delta_c  + \kappa \right) a
-
i
g_0 a (b^\dg + b) \\ \notag
& \ &
+
i g_0 a_p (b^\dg + b)  e^{i \omega_m t}
- i g \sum_{i=1}^N \sigma^i_{ge} 
+ \sqrt{2\kappa} a_{\text{in}} \\
\partial_t b
&=&
-(i\omega_m + \gamma_m) b 
-
i
g_0 
\left( 
a^\dg a 
+  
a^\dg_p a^{\ndg}_p
\right) 
\\ \notag
& \ &
+
i g_0 
\left( 
a^\dg a^{\ndg}_p  e^{i \omega_m  t}
+
a^\dg_p a  e^{-i \omega_m  t}
\right)
+ 
\sqrt{2\gamma_m} \xi ,
\end{eqnarray} 
with $\sigma^i_{ge}:=\ketbra{g_i}{e_i}$.
Here, the in-field operator $a_\text{in}$ includes the steady state average amplitude of the
external field $|\mathcal E_p|$ as well as a fluctuating contribution $\delta a_\text{in}$ characterized by the two-time correlation functions 
\begin{eqnarray}
\ew{\delta a^\ndg_\text{in}(t)\delta a^\dg_\text{in}(t')} &=& \delta(t-t') \\
\ew{\delta a^\dg_\text{in}(t)\delta a^\ndg_\text{in}(t')} &=& 0,
\end{eqnarray}
where we used that the thermal occupation $N_\omega\approx k_B T/\omega$ vanishes for optical frequencies. Here, $k_B$ is the Boltzmann constant and $T$ is the temperature of the bath.
\newline
The Brownian noise $\xi(t)$ affecting the mirror due to coupling to its support satisfies the following correlation function and commutator in the Markovian limit
\begin{eqnarray}
\ew{\xi^\dg(t)\xi(t')} &=& N_m \delta(t-t') \\
\left[ \xi(t),\xi^\dg(t')\right]&=&\delta(t-t').
\end{eqnarray}
The thermal occupation of the mechanical mode is denoted with $N_m$.
\newline
In the limit of an intense pumping field $\mathcal E_p$, the cavity mode $a_p$ can be described by a coherent state $\alpha$ plus an additional fluctuation, and the modified equilibrium position of the resonator is given by $\beta$.
We model this by applying the displacements: $a_p \rightarrow a_p + \alpha$ and $b \rightarrow b + \beta$.
Assuming the steady state condition the equilibrium mean values read
\begin{eqnarray}
\alpha 
&=&
\frac{\mathcal E_p}{\kappa+ig_0 (\beta^* + \beta) + i\Delta_p} \\
\beta 
&=&
\frac{g_0 |\alpha|^2}{i\gamma_m-\omega_m} 
\end{eqnarray}
with $\alpha \gg 1$.
In the steady state limit, the classical ($c$-number) contribution vanishes in the QLEs.
Including only optomechanical interaction strengths of order $\alpha$ and $\alpha^2$
the QLEs for the fluctuation operators read:
\begin{eqnarray*}
\partial_t a_p
&=&
-
(i\tilde \Delta_p+ \kappa ) a_p 
-
i
g_0 \alpha (b^\dg + b)
\\
&\ &
+
i 
g_0 a (\beta^* + \beta) \  e^{-i\omega_m  t} 
+\sqrt{2\kappa} \delta a_\text{in}
\\
\partial_t a
&=&
-
(i\tilde \Delta_c + \kappa) a
- i g \sum_{i=1}^N \sigma^i_{ge} \\
& \ &
+
i g_0 \alpha (b^\dg + b + \beta^* + \beta)  e^{i\omega_m t}
+\sqrt{2\kappa} \delta a_\text{in}
\\
\partial_t b
&=&
-(i\omega_m +\gamma_m) b
-
i
g_0
\left(
\alpha    a^\dg_p
+\alpha^* a^\ndg_p 
\right) \\
& \ &
+i
g_0
\left(
\alpha    a^\dg      e^{i\omega_m t}
+
\alpha^* a^\ndg   e^{-i\omega_m t}
\right) 
+ \sqrt{2\gamma_m} \xi
\end{eqnarray*} 
with the detunings defined by $\tilde \Delta_{n} =\Delta_{n} +g_0 (\beta^* + \beta) $ for $n=\{p,0\}$.
Choosing a rotating frame with respect to: $\tilde \Delta_p a^\dg_p a^\ndg_p + \omega_m b^\dg b$, we require that the pump cavity mode fulfills $\tilde \Delta_p - \omega_m \neq 0$ to apply  the rotating wave approximation with $\alpha g_0 \ll |\tilde \Delta_p - \omega_m| $.
Neglecting terms that are fast oscillating in comparison to the coupling strength, 
the Hamiltonian in leading order reads: 
\begin{eqnarray}
H
\approx
&\ &
\tilde \Delta_c  a^\dg a^\ndg  
+
\Delta_e 
\sum_{i=1}^{N} 
\ketbra{e_i}{e_i}  \\ \notag
&+&
g  \sum_{i=1}^{N} \left(a^\dg \ketbra{g_i}{e_i} + \hc\right)
+
\Omega
\left( \ketbra{e_i}{r_i} + \hc\right)
 \\ \notag
&+&
\left(
G^* \ b^{\dg}   a 
+
G
 b a^\dg
\right) 
 \\ \notag
&+&
\sum_{\substack{i,j=1\\ j>i}}^N \Delta_R \ \ketbra{r_i r_j}{r_i r_j} + \hc ,
\end{eqnarray} 
where $G = -\alpha g_0$. Relabeling $\tilde\Delta_c \rightarrow \Delta_c$ finally leads to the Hamiltonian in Eq.~\eqref{eq:Hamiltonian} in the corresponding rotating frame.

\section{Suppression of the Radiative Decay}
\label{app:radiative_decay_suppression}
\begin{figure}[t!]
\centering
\includegraphics[width=0.5\textwidth]{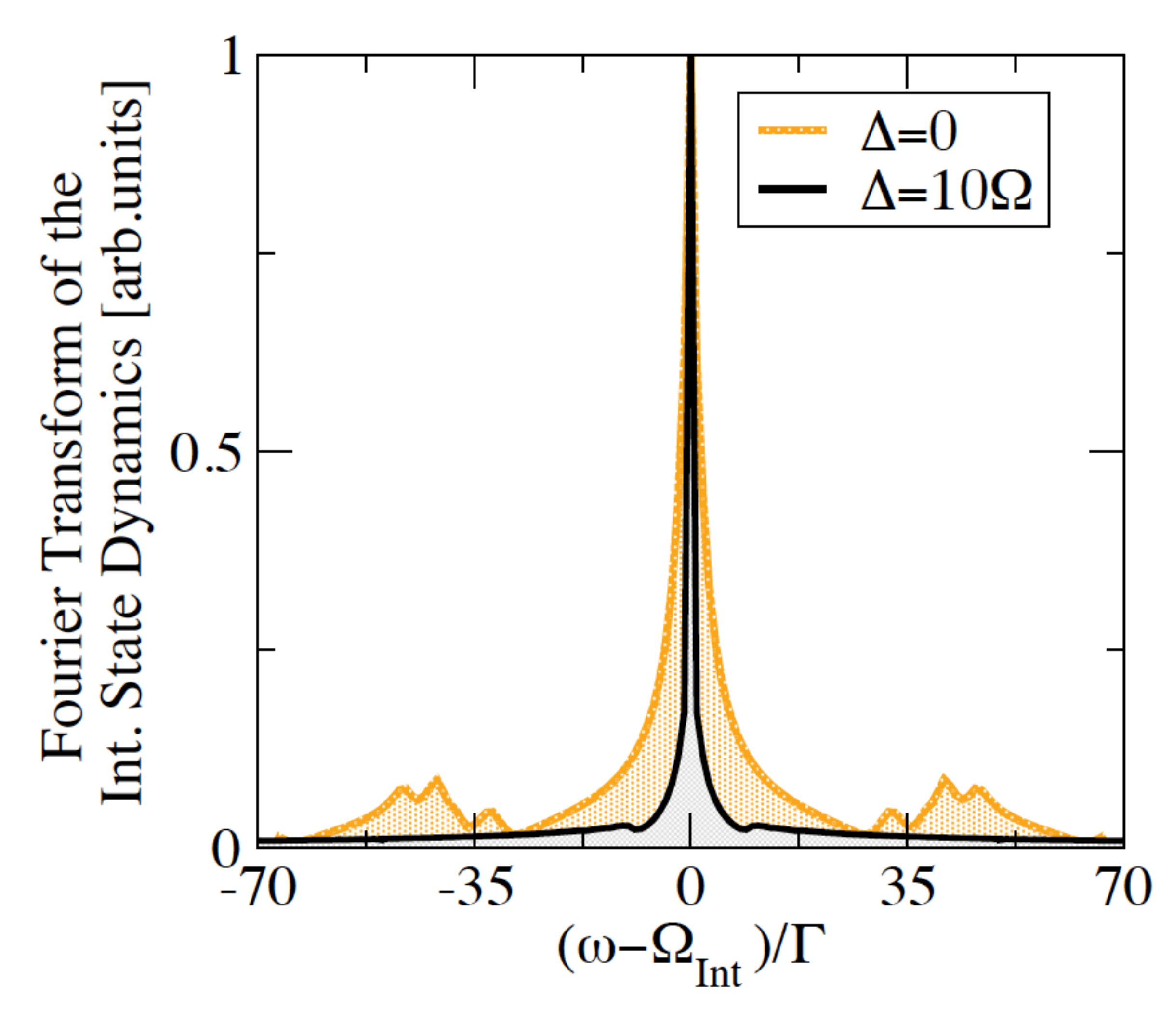}
\caption{The Fourier transform of the intermediate state Rabi oscillations (main contribution) 
are plotted for two different detunings: $\Delta=0$ (orange) and $\Delta=10\Omega$ (black).
Clearly, for a resonant excitation, the Rabi oscillations are stronger damped than in case
of a finite detuning.}
\label{fig:linewidth_compare}
\end{figure}

In the following, we show how the decay of the intermediate excited state is suppressed when we choose a detuned setup.
To illustrate this suppression of the radiative decay, we numerically evaluate a 
semi-classical model governed by the following master equation:
\begin{eqnarray}
\partial_t \rho
&=&
-i \left[ 
H_\Delta, \rho \right] 
+ \Gamma_e
\sum_{i=1}^N
\mathcal D [\ketbra{g_i}{e_i}] \rho
\label{eq:semi_classical} \\
H_\Delta
&=&
\Delta \sum_{i=1}^N \ketbra{e_i}{e_i} 
+ \left( \Omega_{\rm Int} \ketbra{g_i}{e_i}
+ \Omega \ketbra{e_i}{r_i} + \hc \right),
\end{eqnarray}
with one laser (amplitude $\Omega_{\rm Int}$) driving transition between the ground to the intermediate state 
and a second laser (amplitude $\Omega$) driving the transition of the intermediate to Rydberg state.
The Rydberg state is in resonance with the two-photon process of
both lasers, and the intermediate excited state is detuned by $\Delta$.
The radiative decay from the intermediate excited state is chosen
to be $\Gamma_e = \Omega_{\rm Int}/2$.
We numerically evaluate this model for $N=4$ three-level atoms
with a detuning of $\Delta=0$ and $\Delta=10\Omega$
and plot the Fourier transform of the Rabi oscillations of the intermediate
state $\bra{E^1}\rho\ket{E^1}$, defined for the state given in the symmetric
collective basis of Eq.~\eqref{eq:coll_intermed_excited}.
In Fig.~\ref{fig:linewidth_compare}, the corresponding linewidth of the 
intermediate state Rabi oscillations are plotted for both cases.
Clearly, the linewidth with a finite detuning is much smaller than
in the case of a vanishing detuning, approximately four times.
Due to the detuning, the intermediate excited states cannot be
populated fast enough in comparison to the radiative
decay.
As a consequence, the effective radiative decay scales 
not with the number of atoms (if all atoms are excited) 
$N\Gamma_e \rightarrow \Gamma_e$.
Since the radiative decay from the 
intermediate state is the leading source for the population
of non-symmetric states, a large detuning justifies the restriction to the
symmetric subspace, because it suppressed the decay.

\section{Master Equation in the Symmetric Subspace}
\label{app:meq_symmetric_basis}
To rewrite the master equation ~\eqref{eq:full_meq} in the symmetric subspace,
we use the complete set of basis states in case of a single Rydberg excitation 
\begin{eqnarray}
\ket{G} 
&:=& 
\ket{g_1 ... g_N}  \\
\ket{E^j} 
&:=& 
\frac{1}{\sqrt{N_E^j}}
\left(\sum_{i=1}^N \ketbra{e_i}{g_i} \right)^j  \ket{G} \\
\ket{E^j R} 
&:=&  
\frac{1}{\sqrt{N_R^j}}
 \left(\sum_{i=1}^N \ketbra{e_i}{g_i} \right)^{j}
 \left(\sum_{i=1}^N \ketbra{r_i}{g_i}\right)  \ket{G}
\end{eqnarray}
with the normalizations $N^j_E := N!j!/(N-j)!$ and $N^j_R := N\cdot \ N!j!/(N-j)!$ \cite{lukin_prl_rydberg}.
The full Hamiltonian in this basis is given by
\begin{eqnarray} 
\label{eq:collective_hamilonian}
H^N_s 
= 
H^N_{s,0}
+
H^N_{s,\textrm{m-c}}
+
H^N_{s,\textrm{at-l}}
\end{eqnarray}
with the subscript $s$ for symmetric and the superscript $N$ denoting up to N atomic excitations.\footnote{Note, the dipole-dipole interaction is already incorporated 
by considering states with only a single Rydberg excitation.}
The first part of the Hamiltonian governs the free evolution.
We assume the same resonance condition and rotating frame as in Sec.~\ref{sec:adiabatic_elimination}, 
such that the free evolution Hamiltonian reduces to
\begin{eqnarray} 
H^N_{s,0}
& = &  
\Delta_c a^\dg a
+ \Delta_e \sum_{j=1}^N \ j \ 
\ketbra{E^j}{E^j}
+ \Delta_e \sum_{j=1}^{N-1} \ j \ 
\ketbra{E^jR}{E^jR}
\end{eqnarray}
with the detuning of the cavity mode $\Delta_c$ and the detuning of the intermediate
excited state $\Delta_e$. 
The latter scales with the number of excitations
of the intermediate state $j$ in the superatom. 
The interaction of the atomic ensemble with the cavity and external laser reads
\begin{equation} 
\begin{aligned}
H^N_{s,\text{at-l}} 
 =&   
g \sum_{j=1}^N  
\sqrt{j} 
\sqrt{N-(j-1)} \left(
\ a \
 \ketbra{E^{j}}{E^{j-1}}
+\text{h.c.} \right) \\ 
&+g \sum_{j=2}^N 
\sqrt{j} 
\sqrt{N-j} \left(
\ a \
\ketbra{E^{j}R}{E^{j-1}R}
+\hc \right) \notag \\
&+ \Omega \sum_{j=1}^N  \sqrt{j} 
\left( \ketbra{E^{j-1}R}{E^{j}} + \hc \right),
\end{aligned}
\end{equation}
where the first two lines correspond to the atom-cavity interaction and the last line to the atom-laser interaction.
For increasing atomic excitation numbers $j$, the cavity-atom interaction scales 
with $\sqrt{j}$, whereas the collective enhancement factor $\sqrt{N-(j-1)}$ decreases 
correspondingly. The Rabi frequency of the atom-laser interaction is also increased by $\sqrt{j}$ similar to the cavity-atom coupling.

The interaction Hamiltonian between the cavity mode and the membrane is left unchanged such that
$ H^N_{s,\text{m-c}} = H_\text{m-c} $.
Finally, the complete master equation in the symmetric subspace reads:
\begin{eqnarray}
\notag
\partial_t \rho_s
&=&
-i \left[H_s^N, \rho \right]
+ 
\left(
(N_m +1) \ \gamma_m
\mathcal D [b]
+ 
N_m \ \gamma_m 
\mathcal D [b^\dg] \right)
\rho
\\ \notag
& \ &
+ \left(
\kappa
\mathcal D [a] 
+
\sum_{i=0}^{N-1}
\Gamma_r
\mathcal D \left[\ketbra{E^i}{E^iR}\right] 
\right)
\rho
\\ \notag
& \ &
+2 \Gamma_e 
\sum_{i,j=1}^N 
\sqrt{ij(1-\frac{i-1}{N})(1-\frac{j-1}{N})}
\ketbra{E^{i-1}}{E^i}\rho\ketbra{E^j}{E^{j-1}}
\\ \notag
& \ &
+2 \Gamma_e 
\sum_{i=1}^{N-1} 
\sum_{j=1}^{N}
\sqrt{ij(1-\frac{i}{N})(1-\frac{j-1}{N})}
\ketbra{E^{i-1}R}{E^iR}\rho\ketbra{E^j}{E^{j-1}}
\\ \notag
& \ &
+2 \Gamma_e 
\sum_{i=1}^{N} 
\sum_{j=1}^{N-1}
\sqrt{ij(1-\frac{i-1}{N})(1-\frac{j}{N})}
\ketbra{E^{i-1}}{E^i}\rho\ketbra{E^jR}{E^{j-1}R}
\\ \notag
& \ &
+2 \Gamma_e 
\sum_{i=1}^{N-1} 
\sum_{j=1}^{N-1}
\sqrt{ij(1-\frac{i}{N})(1-\frac{j}{N})}
\ketbra{E^{i-1}R}{E^iR}\rho\ketbra{E^jR}{E^{j-1}R}
\\ \notag
& \ &
-\Gamma_e 
\sum_{j=1}^N \left( j \ketbra{E^j}{E^j} + (j-1) \ketbra{E^{j-1}R}{E^{j-1}R}  \right) \rho + \hc \, ,
\end{eqnarray}
where we used that $\ket{E^0}=\ket{G}$.
The important result is that the radiative decay scales only with the number of
excitations and not with the number of atoms.
Note, this master equation is valid for a mainly coherent dynamic, in which
the radiative decay is much smaller than the coherent coupling elements.

\section{Effective Hamiltonian in the Single- and Multi-Excitation Limit}
\label{app:effective_model}

To derive the effective Hamiltonian in the multi-atomic excitation case,
we use the Schr\"odinger equation with the Hamiltonian expressed in terms
of the collective symmetric states:
\begin{eqnarray}
i\partial_t \ket{\Psi} = H_s^N \ket{\Psi} .
\end{eqnarray}
We apply the projector of the relevant sub-space 
\begin{eqnarray}
P^N_\text{m-a} = \ketbra{G,0,N}{G,0,N} + \ketbra{R,0,N\minus1}{R,0,N\minus1}
\end{eqnarray}
with $\ket{R,0,N} := \ket{E^0R,0,N}$
on the wave function:
\begin{eqnarray}
i\partial_t P^N_\text{m-a} \ket{\Psi} &=& P^N_\text{m-a} H_s^N \ket{\Psi} \\  \notag
&=& 
P^N_\text{m-a} H_s^N P^N_\text{m-a} \ket{\Psi}
+
\sum_n P^N_\text{m-a} H_s^N Q_n \ket{\Psi} 
 \\  \notag 
 &=&
P^N_\text{m-a} H_s^N P^N_\text{m-a} \ket{\Psi}
+
P^N_\text{m-a} H_s^N Q_1 \ket{\Psi} ,
\end{eqnarray}
since $P^N_\text{m-a} H_s^N Q_n=0$ for $n > 1$.
As an example, $Q_1$ projects onto the following states
$\left\lbrace \ket{G,1,n-1},\ket{E^1,0,n-1},\ket{E^0R,1,n-2} \right\rbrace$ 
and $Q_2$ consists of the states $\left\lbrace \ket{G,2,n-2},\ket{E^1R,0,n-2},\ket{E^1,1,n-2}, \ket{E^0R,2,n-3} \right\rbrace$.   
To solve this identity, we need furthermore:
\begin{eqnarray}
i\partial_t Q_1 \ket{\Psi} &=& 
Q_1 H_s^N P^N_\text{m-a} \ket{\Psi}
+
\sum_n Q_1 H_s^N Q_n \ket{\Psi} 
 \\  \notag 
 &=&
Q_1 H_s^N P^N_\text{m-a} \ket{\Psi}
+
Q_1 H_s^N Q_1 \ket{\Psi}  
+
Q_1 H_s^N Q_2 \ket{\Psi},
\end{eqnarray}
since also since $Q_1 H_s^N Q_n=0$ for $n > 2$.
It is clear, that the coupling to $Q_2$ leads to a 
higher order contribution in the perturbation theory as second order
Born-Markov.
We assume the detuning large to be enough, that the effective
Hamiltonian can be written as:
\begin{eqnarray}
H^n_\text{eff}
&=&
P^N_\text{m-a} H_s^N P^N_\text{m-a} \\ \notag
& \ &
+
P^N_\text{m-a} H_s^N Q_1 \frac{1}{Q_1 H_s^N Q_1} Q_1 H_s^N P^N_\text{m-a}.
\end{eqnarray}
After evaluating this expression, we obtain the effective Hamiltonian in Eq.~\eqref{eq:n_eff_Hamiltonian}.
As an example, we give the effective master equation in the single-excitation limit, where
the expressions are derived analytically:
\begin{eqnarray}
\notag
\partial_t \rho_\text{s}^1
&=&
-i \left[ H^1_\text{eff}, \rho_\text{s}^1 \right] 
+ \left( 
\Gamma_r
\mathcal D [\ketbra{G,0,0}{R,0,0}]
+
\mathcal D[J_\Gamma ]  
+ 
\mathcal D[J_\kappa ] 
\right) \rho_\text{s}^1
\\ \label{eq:eff_model_single_excitation}
& \ &
+ 
(N_m +1) 
\mathcal D [J_b]
\rho_\text{s}^1
+ 
N_m  
\mathcal D [J^\dg_b]
\rho_\text{s}^1
\end{eqnarray}
with the density matrix $\rho_\text{s}^1$ of the subspace,
in which the membrane and the Rydberg superatom exchange
a single excitation. 
The effective Hamiltonian has now the Jaynes-Cummings form:
\begin{eqnarray}
\notag
H^1_\text{eff}
&=&
-\Delta_g \ketbra{G,0,1}{G,0,1}
-\Delta_r  \ketbra{R,0,0}{R,0,0} \\ 
& \ &
+
G_\text{eff}  
\left( \ketbra{R,0,0}{G,0,1}  + \ketbra{G,0,1}{R,0,0} \right)
\label{eq:eff_h1}
\end{eqnarray}
with the expressions below.
We neglect the contributions due to the finite phonon 
life time and the radiative decay from the Rydberg state, since $\kappa \gg \gamma_m$ 
and $\Gamma_e \gg \Gamma_r$, and we abbreviate $\bar g := \sqrt{N}g$.
The effective coupling between the Rydberg state and the phonon-assisted ground state is derived as:
\begin{eqnarray}
G_\text{eff} &=& \frac{G  \bar g \Omega \left[\Gamma_e\kappa + \bar g^2 - \Delta_e \Delta_c \right]}{(\Gamma_e \kappa +  \bar g^2 - \Delta_c \Delta_e )^2 + ( \Delta_c \Gamma_e +  \Delta_e \kappa )^2}. 
\end{eqnarray}
The dispersive shifts read:
\begin{eqnarray}
\Delta_g &=& 
\frac{G^2 \left[ \Delta_e \bar g^2 - \Delta_c (\Delta_e^2 + \Gamma_e) \right]}{(\Gamma_e \kappa + \bar g^2 - \Delta_c \Delta_e )^2 + ( \Delta_c \Gamma_e +  \Delta_e \kappa )^2} 
\\ \notag
\Delta_r &=& \frac{\Omega^2  \left[ \Delta_c \bar g^2 - \Delta_e (\Delta_c^2 + \kappa) \right]}{(\Gamma_e \kappa +  \bar g^2 - \Delta_c \Delta_e )^2 + ( \Delta_c \Gamma_e +  \Delta_e \kappa )^2} .
\end{eqnarray}
The radiative decay of the intermediate state is transformed into an effective radiative decay of the Rydberg state, and the cavity loss contributes to a phonon loss via:
\begin{eqnarray}
J_\Gamma &=&  \frac{\sqrt{\Gamma_e} \ket{G,0,0} \left( i \bar g G \bra{G,0,1} - \Omega (i\Delta_c + \kappa) \bra{R,0,0} \right) }{\sqrt{(\Gamma_e \kappa + \bar g^2 - \Delta_c \Delta_e )^2 + ( \Delta_c \Gamma_e +  \Delta_e \kappa )^2}} \\ \notag
J_\kappa &=& \frac{\sqrt{\kappa}  \ket{G,0,0} \left( i \bar g \Omega \bra{R,0,0} - G (i\Delta_e + \Gamma_e) \bra{G,0,1} \right) }{\sqrt{(\Gamma_e \kappa + \bar g^2 - \Delta_c \Delta_e )^2 + ( \Delta_c \Gamma_e +  \Delta_e \kappa )^2}}.
\end{eqnarray}

\section{Long-Distance Coupling}
\label{app:long_distance_model}
In the following we outline the derivation of an effective master equation for the coupling between the membrane and the superatom. The derivation is done in three steps: first, we formulate the full Hamiltonian for the long-distance setup, then we eliminate the mediating coupling field, and finally we transform the resulting master equation into the collective basis, as described in Sec.~\ref{sec:cohdynSA}.

The full Hamiltonian for the problem reads:
\begin{align}
\label{eq:Htot}
\tilde{H}&=\tilde{H}_{0}+\tilde{H}_{\textnormal{m-f}}+\tilde{H}_{\textnormal{at-f}}\,.
\end{align}
with $\tilde{H}_0$ defined in Eq.~\ref{eq:H0tilde}, $\tilde{H}_{\textnormal{m-f}}$ defined in \cite{vogell_pra_cavity_enhancement} and $\tilde{H}_{\textnormal{at-f}}$ given by
\begin{align}
H_{\textnormal{at-f}}&= \hbar \sum_j \left(\Omega_L^m \sigma_{eg}^j \int \! \text{d}\omega \, c_\omega \sin(k^m_L z_j) + \Omega_L e^{-i \omega_L t} \sigma_{er}^j +\hc \right),
\label{eq:Hintatf}
\end{align}
where $ \ketbra{e_i}{g_i}=\sigma^i_{eg}$, Rabi frequencies $\Omega_L^m$ and $\Omega_L$. In contrast to the cavity-mediate case, where we coupled to a single cavity mode, we assume here a continuum of field modes $c_\omega$ centered around the coupling
laser frequency $\omega_L^m$, that mediate the interaction. 
The membrane-light field interaction can be linearized around a strong laser drive at frequency $\omega_L^m$ such that it is given by
\begin{eqnarray}
\tilde{H}^\textnormal{lin}_\textnormal{m-f}&= \hbar
g_m \left(
b \int \! d\omega   \,
c^\dg_\omega + \hc \right),
\end{eqnarray}
where $g_m$ is the laser enhanced coupling element derived in \cite{vogell_pra_cavity_enhancement}.\\

In order to eliminate the intermediate excited state, we choose the frequency of the driving field to be off-resonant with the transition of the ground to intermediate state, i.e. $\Delta_e=\omega_e - \omega_L^m$,
 to suppress a resonant driving of the atomic
ensemble. The resonance condition reads $\omega_r = \omega_L + \omega_L^m+\omega_m$ as in Sec.~\ref{sec:cavity_mediated}.
Due to $\Delta_e \gg \Omega^m_{L},\Omega$ we can then eliminate the intermediate excited state. We further need assume the same delay throughout the atomic ensemble, i.e. $\tau_j = k z_j/c \approx k {\bar z}_j /c = \tau$ with an average position ${\bar z}_j$.

We further need to eliminate the mediating quantized laser field, which is in a very similar way done in \cite{vogell_unpublished}, and it uses the framework of the quantum stochastic Schr\"odinger equation, cf.
\cite{hammerer_pra_optical_lattices_om,gardiner-book}.
The elimination finally leads to a master equation for a cascaded quantum system,
where the dynamics cannot longer be described by a unique Hamiltonian. The coherent dynamics is governed by the induced Hamiltonian
\begin{eqnarray}
H_\text{ ind} 
=&
-&\hbar
\sum_j \Delta_\textnormal{at}^j \ \sin(2k_L^m \bar{z}_j) \ \sigma^j_{rr} \notag \\
&+&
\hbar 
\sum_j
\left( \tilde{G}_\textnormal{eff}^j \ \cos(k_L^m \bar{z}_j ) \ b^\dg  \sigma^j_{gr} + \hc \right), 
\end{eqnarray}
where $\tilde{G}^j_\textnormal{eff}:=g_m g_{\text at}^j$ is the effective coupling and $\Delta_\textnormal{at}^j:=(g^j_{\text at})^2$ a dispersive shift.
The full master equation reads
\begin{eqnarray}
\dot \rho
=& \ &
-\frac{i}{\hbar} \left[H_\text{ ind},\rho\right] 
+ 
\frac{\gamma_m^\textnormal{diff}}{2} \mathcal{D}\left[b^\dg \right]  \rho \\ \notag
& \ & 
+4 \sum_{i,j=1}^N
g_{\text at}^i g_{\text at}^j \sin(k_L^m \bar{z}_i) \sin(k_L^m \bar{z}_j)
\left( 
\sigma_{gr}^i \rho \sigma_{rg}^j - \frac{\delta_{ij} }{2} \sigma^i_{rr} \rho - \frac{\delta_{ij} }{2} \rho \sigma^i_{rr} 
\right) \\ \notag
& \ & 
+2 \sum_{j=1}^N \left(
\tilde{G}_\textnormal{eff}^j \sin(k_L^m \bar{z}_j) 
\left( 
\sigma_{gr}^i \rho b^\dg  
- \frac{1}{2}  b^\dg \sigma^i_{gr}\rho - \frac{1}{2} \rho b^\dg \sigma^i_{gr}  \right)
+ \hc \right),
\end{eqnarray}
where $\delta_{ij}$ is the Kronecker delta and $\gamma_m^\textnormal{diff}:=2 g_m^2$
is the membrane-light field diffusion as in \cite{vogell_pra_cavity_enhancement}.

This master equation is still in the microscopic single particle picture of the atomic ensemble. 
In order to rewrite the master equation in the collective basis for a single Rydberg excitation and thereby reducing the atomic ensemble to an effective two-level atom, 
we assume, that the atoms in the ensemble couple equally strong to the light field $g_\text{at}^i = g_\text{at}$ and remain at constant position $\bar{z}_j = \bar{z}$. Further, we then have $\tilde{G}^j_\textnormal{eff}=\tilde{G}_\textnormal{eff}$ and $\Delta_\textnormal{at}^j=\Delta_\textnormal{at}$.

Under the condition that only a single Rydberg excitation is possible due to a Rydberg blockade radius, we can restrict the Hilbert space to the states: $\ket{G}$ and $\ket{E^0R}=:\ket{R}$. 
Using this we conclude with the Hamiltonian in Eq.~\eqref{eq:Hcollind} and the associated master equation \eqref{eq:HcollindMEQ}.


\end{appendix}

\noindent
\section*{References}

\bibliographystyle{unsrt}

\bibliography{membrane_superatom}

\end{document}